\documentclass[a4paper,10pt,twocolumn]{cpc-hepnp}

\usepackage{multicol}
\usepackage{graphicx}
\usepackage{booktabs}
\usepackage{amssymb,bm,mathrsfs,bbm,amscd}
\usepackage[tbtags]{amsmath}
\usepackage{lastpage}
\usepackage{CJK}

\usepackage{hyperref}

\usepackage{float}

\usepackage{umoline}
\usepackage{graphicx}
\usepackage{textcomp}

\usepackage{tabulary}

\usepackage{amsmath}

\usepackage{gensymb}
\usepackage{url}

\usepackage{xcolor}

\usepackage{lineno}

\usepackage{float}

\usepackage{booktabs}
\usepackage{siunitx}

\usepackage{chngcntr}

\RequirePackage[numbers,sort&compress]{natbib}

\begin{document}

\fancyhead[c]{\small Submitted to Chinese Physics C~~~} \fancyfoot[C]{\small \thepage}

\footnotetext[0]{Received ---}

\title{Study of Monte Carlo event generators for proton-proton collisions at LHC energies in the forward region\thanks{Supported through the following projects: Proiect IDEI, contract number 56/07.10.2011 and the IFA project Rom\^{a}nia-CERN, contract number 7/16.03.2016 (LHCb-Ro)}}

\author{%
      Alexandru C\u{a}t\u{a}lin ENE$^{1,2;1)}$\email{alexandru.catalin.ene@cern.ch}%
\quad Alexandru JIPA$^{2}$%
\quad Lavinia-Elena GIUBEGA$^{1}$
}

\maketitle

\address{%
$^1$ Department of Elementary Particles Physics, Horia Hulubei National Institute for R\&D in Physics and Nuclear Engineering (IFIN-HH), Reactorului 30, RO-077125, P.O.B. MG-6, M\u{a}gurele-Bucharest, Romania\\
$^2$ Faculty of Physics of the University of Bucharest, Atomi\c{s}tilor 405, RO-077125, P.O.B. MG-11, M\u{a}gurele-Bucharest, Romania\\
}



\begin{abstract}
In this paper we present a comparative study between PYTHIA, EPOS, QGSJET, and SIBYLL generators. The global event observables considered are the charged energy flow, charged-particle distributions, charged-hadron production ratios and $V^{0}$ ratios. The study is performed in the LHCb and TOTEM fiducial phase-spaces on minimum bias simulated data samples for \emph{pp} collisions at $\sqrt{s} = 7$ TeV using reference measurements from the aforementioned experiments. In the majority of cases, the measurements are within a band defined by the most extreme predictions. The observed differences between the predictions and the measurements seem to be, in most part, caused by extrapolation from the central pseudorapidity region ($|\eta| \leq$ 2.5), in which the generators were mainly tuned.
\end{abstract}

%\keywords{Cosmic rays, pp collisions, LHCb, Monte Carlo generators, collider, PYTHIA, EPOS, QGSJET, SIBYLL}

\begin{multicols}{2}

\section{Introduction}

One of the most important sources of information concerning elementary particle physics is the study of high energy cosmic rays. Up until the advent of powerful particle accelerators in the 1950s, the only source of high energy particles were the cosmic rays. The cosmic ray spectrum reaches energies of the order of 10\textsuperscript{20} eV \cite{1}, whilst the most powerful collider to date, the Large Hadron Collider, reaches energies of 13 TeV in the center of mass frame or about 10\textsuperscript{17} eV fixed target equivalent. So, there are two independent sources of information for \emph{pp} collisions at the same energy scale. Combining the two helps create a better picture of the phenomena that take place in such collisions. Although the cross-section of hard interactions is considerable at these energy scales, the soft interaction part is still large. As soft processes imply non-perturbative QCD, we rely on phenomenological models and effective theories for predictions. Hadronic interactions generators have been developed for the description of the physics at the aforementioned energy scales, with an emphasis on either cosmic rays or collider physics. In recent years, cosmic rays generators have been extensively tuned to collider physics measurements, especially in the context of the newly available data from LHC. In this paper we compare the predictions obtained EPOS LHC \cite{3}, QGSJETII-04 \cite{6} and SIBYLL 2.3 \cite{7} generators included in the CRMC package \cite{16} and the widely used event generator for LHC physics, PYTHIA (versions 8.186 \cite{37} and 8.219 \cite{8}) for \emph{pp} interactions at $\sqrt{s}$ = 7 TeV with measurements from the LHCb and TOTEM experiments. The generators studied are all tuned using various observables measured at LHC experiments. Predictions obtained with PYTHIA 8.186 using the non-LHC tune 2M are also shown for reference. Throughout this paper we are referring to measurements/tunes performed in the ``central" and ``forward" regions defined with respect to the pseudorapidity of the particles. The central pseudorapidity region is defined as $|\eta| \leq 2.5$, corresponding to the ATLAS, ALICE and CMS acceptances \cite{46,47,48}, and the forward pseudorapidity region as $\eta \geq 2.5$, corresponding to the LHCb ($2\leq \eta \leq 5$) and TOTEM ($3.1 \leq |\eta| \leq 6.5$) acceptances \cite{49,50}.

\section{The Monte Carlo event generators}

\subsection{General description}

The generators used for this study are PYTHIA, a collider physics generator, EPOS, QGSJET and SIBYLL, which are cosmic ray collisions generators. They can be split in three categories according to the models on which they are based. PYTHIA is a parton based generator and it simulates parton interactions and parton showers, the hadronization being treated using the Lund string fragmentation model \cite{2,29}. Another category would be the one of the generators based on the Regge theory such as QGSJET and SIBYLL. These models treat soft and semihard interactions as Pomeron exchanges (``soft" and ``semihard" Pomerons), but also mix perturbative methods into the treatment of hard interactions \cite{4,29}. EPOS is part of a distinct category in which the parton based description is mixed with aspects from the Regge theory \cite{29}. The focus of the study is on minimum bias physics measurements and the generators used, especially the cosmic ray ones, are developed for the description of such observables. The selection of these particular generators is justified by their varied usage and basic assumptions, while at the same time sharing similarities and being tuned to LHC data, as it will be disccused below.

PYTHIA is one of the most used Monte Carlo event generator for collider physics with an emphasis on \emph{pp} interactions. It is mainly based on Leading Order (LO) QCD, having implemented LO matrix elements and usualy using LO PDF sets (NLO PDF sets also available) \cite{8, 33, 34}. The main event in a \emph{pp} collision (internally called ``hard process") can be represented by a plethora of processes like elastic and diffractive (described using Pomerons) \cite{2, 8, 36}, soft and hard QCD processes, electroweak processes, top quark production etc. The generator also implements parton showers (Initial State Radiation, ISR, and Final State Radiation, FSR) in Leading Log (LL) approximation with matching and merging methods between them and the hard processes \cite{8, 33}. Given that the colliding hadrons have a complex partonic structure, other partonic interactions aside from the main event are expected. These are called multiparton interactions (MPI) and are usually soft in nature, but the momentum transfer can also reach the hard interaction energy scale. PYTHIA implements a description of both types and also of the beam remnants which form after the extraction of MPI initiator partons \cite{8}. The hadronization mechanism is based on the Lund string fragmentation model \cite{8}.

The Parton-Based Gribov-Regge Theory is an effective field theory using concepts from QCD in which the elementary interactions between the constituent partons of nucleons/nuclei proceed via exchanges of parameterised objects called Pomerons which have the quantum numbers of the vacuum \cite{22, 35}. In this theory the elementary collisions are treated as a sum of soft, semihard and hard contributions. If one considers a cutoff value of the momentum transfer squared of $Q_{0}^{2}\sim 1$ GeV\textsuperscript{2}, below which perturbative QCD calculations can no longer be done, then the soft contribution (non-perturbative) is represented by processes with $Q^{2}<Q_{0}^{2}$ and the hard contribution (perturbative) by processes with $Q^{2} > Q_{0}^{2}$. The processes in which sea partons with $x\ll1$ (Bj\"{o}rken \emph{x}) are involved are called semihard and are represented by a parton ladder with soft Pomeron ends \cite{22}.

The generator EPOS is based on the effective theory described above \cite{3}. EPOS is an acronym for \textbf{E}nergy conserving quantum mechanical approach, based on \textbf{P}artons, parton ladders, strings, \textbf{O}ff-shell remnants, and \textbf{S}plitting of parton ladders \cite{30}. In EPOS the interaction of the two beam particles is described by means of Pomeron exchanges. As discussed above, these Pomerons can be soft, semihard or hard. A soft Pomeron can be viewed from a phenomenological standpoint as two parton ladders (or cut Pomeron) connected to the remnants by two color singlets (legs) from the parton sea \cite{21}. A cut Pomeron can be viewed as two strings which fragment to create hadrons. The flavours of the string ends need to be compensated within the remnants. Thus, particle production in EPOS comes from two sources, namely cut Pomerons and the decay of remnants \cite{21}. Through a recent development (from EPOS 1.99 onwards), EPOS is now a core-corona model. The core represents a region with a high density of string segments that is larger than some critical density for which the hadronization is treated collectively and the corona is the region with a lower density of string segments for which the hadronization is treated non-collectively. The strings from the core region form clusters which expand collectively. This expansion has two components, namely radial and longitudinal flow. Through this core-corona approach, EPOS takes into account effects not accounted for in other HEP models \cite{3}. In EPOS, in the case of multiple scatterings (multi-Pomeron exchanges) the energy scales of the individual scatterings are taken into account when calculating the respective cross-sections, while in other models based on the Gribov-Regge Theory this is not the case. This leads to a consistent treatment of both exclusive particle production and cross-section calculation, taking energy conservation into account in both cases \cite{21,22}. The multiplicity and inelastic cross section predictions of the model are directly influenced by energy momentum sharing and beam remnant treatment \cite{21}.

The elementary scatterings in QGSJET are also treated as Pomeron exchanges \cite{4}. QGSJET is based on the Quark-Gluon string model, which is in turn based on the Gribov-Regge model \cite{31}. In this model the Pomeron exchange can be viewed as an exchange of a non-perturbative gluon pair. Each of the colliding protons can be considered as being a system of a quark and a diquark with opposite transverse momenta. The quark from the first proton exchanges a non-perturbative gluon with the diquark from the second proton and viceversa, thus creating two quark-gluon strings which will decay according to fragmentation functions to create hadrons \cite{32}. In a similar manner to EPOS, the soft (non-perturbative) and hard (perturbative) contributions are separated by a cutoff value of $Q_{0}^{2}$. In QGSJET a Pomeron is actually a sum of two contributions: a ``soft" Pomeron one and a ``semi-hard" Pomeron contribution. The soft part represents a purely non-perturbative parton cascade, while the ``semi-hard" Pomeron can be viewed as two ``soft" Pomerons connected by a parton ladder \cite{45}. At very high energies as those at the LHC and/or small impact parameters, the semi-hard contribution dominates and so it is crucial to take it into account \cite{4,31}. In these high energy collisions large numbers of parton-parton interactions occur, the resulting cascades interacting with one another (Pomeron-Pomeron interactions) and thus their evolution is no longer indepedent, but correlated. QGSJET-II takes into account these non-linear effects which are computed with enhanced Pomeron diagrams \cite{4,31}.

SIBYLL is based on the dual parton model (DPM), using the minijet model for hard interactions and the Lund string fragmentation model for hadronization \cite{5,23}. Similarly to both EPOS and QGJSET, soft and hard interactions are separated by a transverse momentum scale cutoff value. The soft interactions are treated using the dual parton model (DPM) in which the nucleon is treated as consisting of a quark and a diquark, and similar to the Quark-Gluon string model described above, a quark (diquark) from the projectile combines with the diquark (quark) from the target to form two strings which are fragmented separately using the Lund string fragmentation model. In SIBYLL 1.7 the cutoff value was set to $p_{T}^{min}=\sqrt{5}$ GeV, but from version 2.1 onwards it was changed to a function of the collision energy which for $\sqrt{s}=7$ TeV returns $p_{T}^{min}\approx3,87$ GeV \cite{5}.

\subsection{Versions used in the study}

The default tune for PYTHIA 8.186 is Tune 4C with the CTEQ6L1, LO PDF set as the default one \cite{8,9}. Tune 4C (default from version 8.150 onwards \cite{10}) is obtained starting from Tune 2C for which Tevatron data have been used, by varying MPI and colour reconnection parameters to fit the measurements for minimum bias (MB) and underlying event (UE) observables from ALICE and ATLAS experiments at various collision energies (0.90, 2.36 and 7 TeV). The observables used are, for example: charged multiplicity and rapidity distributions, transverse momentum distributions, mean transverse momentum as a function of charged multiplicity distributions, transverse momentum sum densities etc. Tune 2M is obtained in a similar manner to 2C, using measurements from the CDF experiment at Tevatron, but uses the modified PDF set MRST LO** instead of the CTEQ6L1, LO PDF set \cite{39}. From here on, PYTHIA 8.186 with Tune 2M will be refered to as PYTHIA 8.1 2M. 

PYTHIA 8.219 has the Monash 2013 tune as it's default (with the NNPDF3.3 QCD+QED LO PDF set) \cite{8,10}. The Monash 2013 tune has been created for a better description of minimum bias and underlying event observables. Similar observables as for the previous tune have been used, with measurements from ATLAS and CMS experiments, and the charged pseudorapidity distribution from TOTEM in the forward region. The flavour-selection parameters of the string fragmentation model have been re-tuned using a combination of data from PDG and from the LEP experiments, resulting in an overall increase of about 10\% in strangeness production and a similar decrease of the production of vector mesons. The kaon yields have clearly improved with respect to CMS measurements and the ones of hyperons are also slightly improved. The minimum bias charged multiplicity has also increased by about 10\% in the forward region \cite{19}.

EPOS LHC's fundamental parameters are tuned to cross-section measurements from the TOTEM experiment at $\sqrt{s} = 7$ TeV, leading to a highly improved description of charged multiplicity (compared to EPOS 1.99). In EPOS LHC the radial flow calculations are corrected. This correction affects the high multiplicity region, again leading to a highly improved description of this observable in this particular region. In EPOS 1.99 the baryon-antibaryon pair and strangeness production were largely overestimated in high energy collisions. This issue was corrected in EPOS LHC and by using the same string fragmentation parameters as for $e^{+}e^{-}$ collisions, kaon/pion and proton/pion ratio measurements from CMS at $\sqrt{s}=$ 7 TeV are reasonably well described \cite{3}. The statistical particle production mechanism from the core affects strangeness production by removing its suppression. This leads to a good description of strange baryon yield measurements from CMS at $\sqrt{s}=$ 7 TeV as shown in Figure 10 from \cite{3}. The radial flow parameters are tuned using charged-particle transverse momentum distributions (for minimum bias \emph{pp} collisions) obtained at the ATLAS experiment at $\sqrt{s}=$ 0.9 and 7 TeV. This leads to a very good agreement with experimental transverse momentum distributions of identified particles \cite{3}.

QGSJETII-04 distinguishes itself from the previous version, QGSJETII-03, by taking into account all significant enhanced Pomeron diagram contributions, including Pomeron loops, and the tuning to new LHC data \cite{40}. As QGSJET is used for high energy cosmic rays studies, the current version of the generator has been tuned to LHC measurements for observables to which the extensive air shower (EAS) muon content is sensitive. Examples of such observables are: charged particle multiplicities and densities, anti-proton and strange particle yields etc. QGSJETII-03 predicts a steeper increase in multiplicity in pseudorapidity plots from $\sqrt{s}=$ 0.9 to 7 TeV than what is observed in ATLAS measurements for these collision energies. As a consequence, the $Q_{0}^{2}$ separation scale between soft and hard interactions has been increased from 2.5 GeV$^{2}$ to 3.0 GeV$^{2}$. For a better description of ALICE measurements of the antiproton transverse momentum spectrum at $\sqrt{s}=$ 0.9 TeV, the anti-nucleon yield was slightly reduced and the hadronization parameters have been modified as to enlarge the average transverse momentum of the anti-nucleons. The strangeness production has been enhanced to better describe $K_{S}^{0}$ and $\Lambda$ rapidity distributions measured at CMS for $\sqrt{s}=$ 0.9 TeV and 7 TeV \emph{pp} collisions. Another major tuning is done using inelastic cross section measurements at $\sqrt{s}=$ 7 TeV from the TOTEM experiment \cite{41}.

SIBYLL is a relatively simpler model and emphasis is put on describing observables on which the evolution of extensive air showers depends, like energy flow and particle production in the forward region \cite{42}. In SIBYLL 2.3 soft gluons can be exchanged between sea quarks or sea and valence quarks also. A new feature in version 2.3 is the beam remnant treatment which is similar to that of QGSJET. This new treatment allows the particle production in the forward region to be tuned without modifying the string fragmentation parameters. A major tuning procedure has been done for the description of leading particle measurements from the NA22 and NA49 experiments \cite{7}. SIBYLL 2.3 has also been tuned using measurements from $\sqrt{s}= 7$ TeV \emph{pp} collisions at LHC experiments namely, the inelastic cross section from TOTEM, average antiproton multiplicities and charged particle differential cross sections as a function of transverse momentum obtained at CMS. The SIBYLL 2.1 version was tuned using Tevatron data and it describes, for example, charged pseudorapidity density measurements reasonably well, even the ones from CMS at $\sqrt{s}=$ 7 TeV, as one can see in Figure 4 from \cite{43}. At the same time SIBYLL 2.1 overestimates the inelastic cross section measurements at high collision energies (beyond 1 TeV), leading to the tuning of version 2.3 with the $\sigma_{pp}^{inel}$ measurements at $\sqrt{s}=$ 7 TeV from TOTEM. The antiproton multiplicities measured in fixed target experiments at low collision energies seem to be reasonably well described by version 2.1, but the measurements obtained at the CMS experiment for various collision energies are largely underestimated. To correct this effect in SIBYLL 2.3, a different value of the quark/diquark production probability, $P_{q/qq}$,  has been assigned for the fragmentation of minijets than for all the other fragmentation processes. The value of $P_{q/qq}$ in SIBYLL 2.1 was fixed to 0.04 for all processes. SIBYLL 2.3 uses the same effective parton density function as the previous version, but the quark and gluon contributions are obtained from the same parametrizations used to calculate the minijet cross section. This leads to a steeper parton distribution function at low Bj\"orken \emph{x} which combined with the correction of the definition of $p_{T}^{min}$, leads in turn to a better description of the measurements for charged particle cross sections as a function of transverse momentum obtained at CMS in the $2\leq p_{T} \leq5$ GeV/c range. Also, a charm hadron production model was implemented in version 2.3 \cite{43}.

\section{Data generation and analysis strategy}

Samples of $10^{6}$ inelastic minimum bias \emph{pp} events at $\sqrt{s}$ = 7 TeV were generated for each generator. For all generators a stable particle definition of $c\tau\ge$ 3 m was used, where $\tau$ is the mean proper lifetime of the particle species.

This study treats five distinct aspects: charged energy flow, charged-particle distributions, charged-hadron production ratios and $V^{0}$ ratios.

Charged energy flow is computed as the total energy of stable charged particles ($p$, $\bar{p}$, $K^{\pm}$, $\pi^{\pm}$, $\mu^{\pm}$ and $e^{\pm}$) in the interval 1.9 $\le\eta\le$ 4.9 (10 bins of $\Delta\eta=$ 0.3), divided by the width of the pseudorapidity bin and normalised to the number of visible inelastic \emph{pp} interactions $N_{int}$ or:

\begin{equation}
\frac{1}{N_{int}}\frac{dE_{total}}{d\eta}=\frac{1}{\Delta\eta}\Bigg(\frac{1}{N_{int}}\sum_{i=1}^{N_{part},\eta} E_{i,\eta}\Bigg),
\end{equation}

\noindent
where $N_{part},\eta$ is the number of stable charged particles (as defined above) in a $\Delta\eta=$ 0.3 bin and $E_{i,\eta}$ is the energy of the particles from the respective bin (see \cite{11}).

There are four event classes considered for the charged energy flow: inclusive minimum bias events, hard scattering events, diffractive enriched events and non-diffractive enriched events. The inclusive minimum bias events are required to have at least one charged particle in the range: 1.9 $\le\eta\le$ 4.9. The hard scattering events require at least one charged particle with $p_{T}\ge3$ GeV/c in the aforementioned range. Diffractive enriched events require that no particles are generated in the pseudorapidity range of $-3.5<\eta<-1.5$ and non-diffractive enriched events require at least one particle in this range. These event class definitions are compatible with the ones from \cite{11} from which the LHCb reference measurements were taken.

The purity of the diffractive enriched and non-diffractive enriched events samples have been studied for both versions of PYTHIA (as the generator has readily accessible event type information) and are about 94\% and 92\%, respectively. In Figure \ref{fa1}, the transverse momentum scale distributions of the hardest parton collision from hard and soft (non-hard and non-diffractive) events, obtained with PYTHIA 8.186, are shown. As can be seen, the peaks are reasonably well separated with $\mu\approx8.7$ GeV/c, $\sigma\approx4.5$ GeV/c, for hard events and $\mu\approx4.2$ GeV/c, $\sigma\approx3.2$ GeV/, for soft events. The fraction of events that pass both the hard and diffractive enriched event class conditions are negligible.

The values of the number of visible events for the different event classes are given in Table \ref{t1}.

The transverse momentum, pseudorapidity and multiplicity distributions of charged stable particles (\emph{p}, $\pi$, $K$, \emph{e}, $\mu$)  are presented in Figures \ref{f3}-\ref{f6}. The distributions were scaled with the number of visible events from the sample. The visible events are required to contain a minimum of one charged-particle satisfying the criteria listed below:

\begin{itemize}
\renewcommand{\labelitemi}{$\bullet$}
  \item Figure \ref{f3}: $2 <\eta < 4.8$, $p\ge2$ GeV/c and $p_{T}>0.2$ GeV/c \cite{14}.
  \item Figure \ref{f4}: $2 <\eta < 4.5$ \cite{25}.
  \item Figure \ref{f5}: $2.5 <\eta < 4.5$ and $p_{T}>1$ GeV/c. These events will be called ``hard" \cite{25}.
  \item Figure \ref{f6}: $5.3 <\eta < 6.5$ and $p_{T}>40$ MeV/c \cite{27}.
\end{itemize}

\vspace{-0.4 cm}

\begin{center}
    \includegraphics[trim={0.7cm 1.8cm 0.7cm 2cm},clip,width=0.45\textwidth]{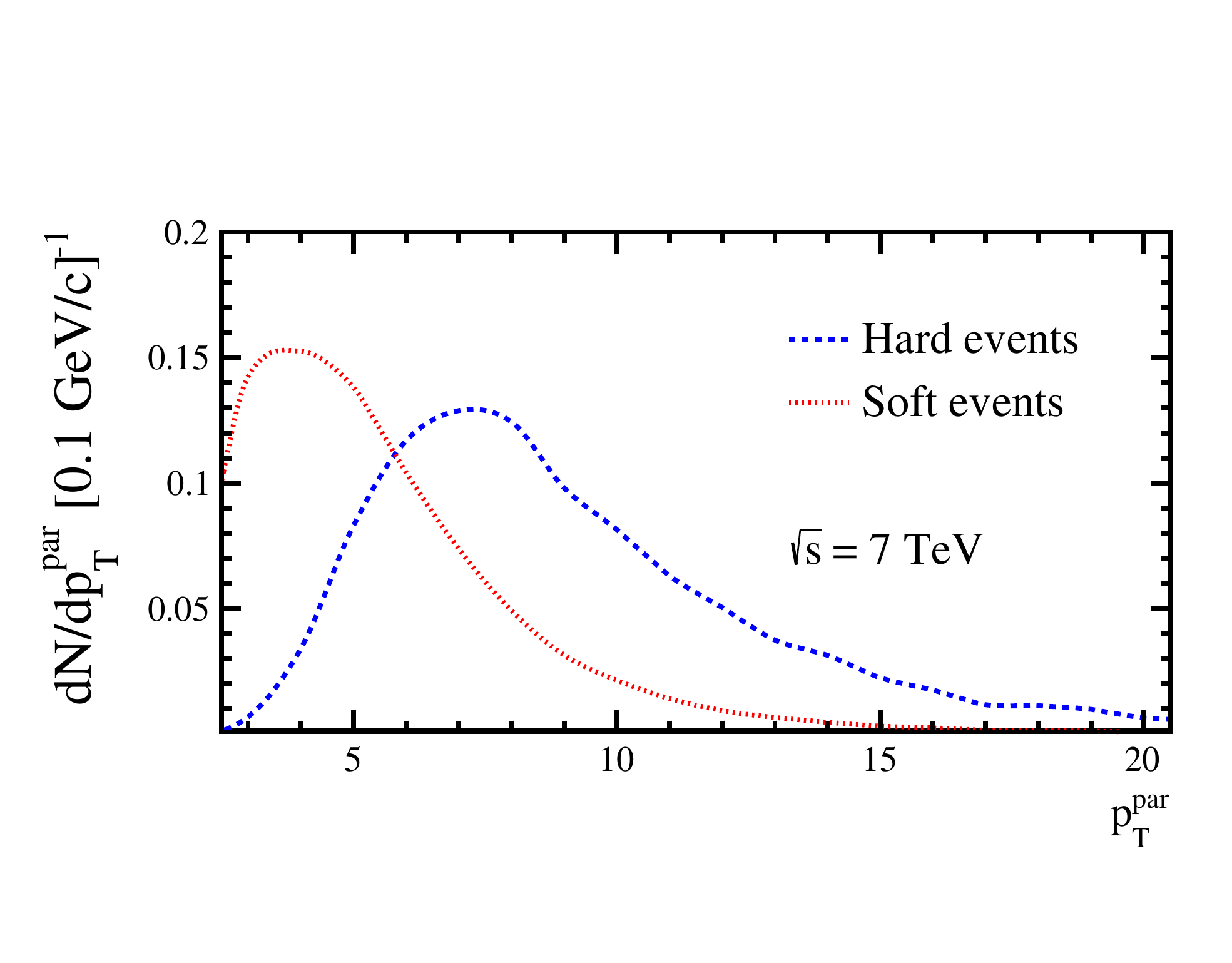}

 \figcaption{\label{fa1} Transverse momentum scale of the hardest subprocess obtained with PYTHIA 8.186 for hard and soft events. The distributions were normalized with the number of visible events for each event class.}

\end{center}

The numbers of minimum bias and hard events with a minimum of one charged particle in the range $2 <\eta < 4.5$ are given in Table \ref{t2}.

For all of the distributions mentioned above pull plots of ($x_{gen}-x_{exp})/\sigma_{exp}$ have been drawn.

A particle is defined as prompt if the sum of it's ancestors' mean proper lifetimes is less than 10 ps as in \cite{14,25,27}.

The prompt charged-hadron production ratios as a function of pseudorapidity are shown in Figures \ref{f7}-\ref{f9} and are the following: $\bar{p}/p$, $\pi^{-}/\pi^{+}$, $K^{-}/K^{+}$, $(K^{+}+K^{-})/(\pi^{+}+\pi^{-})$, $(p+\bar{p})/(K^{+}+K^{-})$ and $(p+\bar{p})/(\pi^{+}+\pi^{-})$. These ratios are computed in the phase-space defined by $2.5\le\eta\le4.5$ and $p\ge5$ GeV/c and three transverse momentum intervals, namely $p_{T}<$ 0.8 GeV/c, 0.8 $\le p_{T}<$ 1.2 GeV/c and $p_{T}\ge$ 1.2 GeV/c \cite{12}. 

The prompt $V^{0}$ particle ratios $\bar{\Lambda}/\Lambda$ and $\bar{\Lambda}/K_{S}^{0}$ as a function of rapidity are shown in Figure \ref{f10}. The ratios are computed in the phase-space defined by $2\le y \le4.5 $ and three $p_{T}$ intervals: 0.15 $<p_{T}<$ 0.65 GeV/c, 0.65 $<p_{T}<$ 1.00 GeV/c and 1.00 $<p_{T}<$ 2.50 GeV/c. Figures \ref{f11}-\ref{f12} show the prompt $V^{0}$ particle ratios as a function of rapidity and as a function of transverse momentum in the $2\le y \le4.5 $ rapidity interval and the full $p_{T}$ interval 0.15 $<p_{T}<$ 2.50 GeV/c \cite{13}. 

The statistical uncertainties of the MC predictions are negligible, reaching a maximum value of about 3 \% in the least populated bins at the edges of the considered phase-space regions, while for the rest of the bins the uncertainties are of the order of 0.1 \%.

The sources of the reference measurements used in the plots are given at the end of the captions.

\section{Results and discussion}

\begin{figure}[t] 

  \begin{minipage}[b]{0.5\linewidth} 
    \centering
    \includegraphics[trim={0.8cm 0.cm 0.7cm 0.5cm},clip,width=0.9\textwidth]{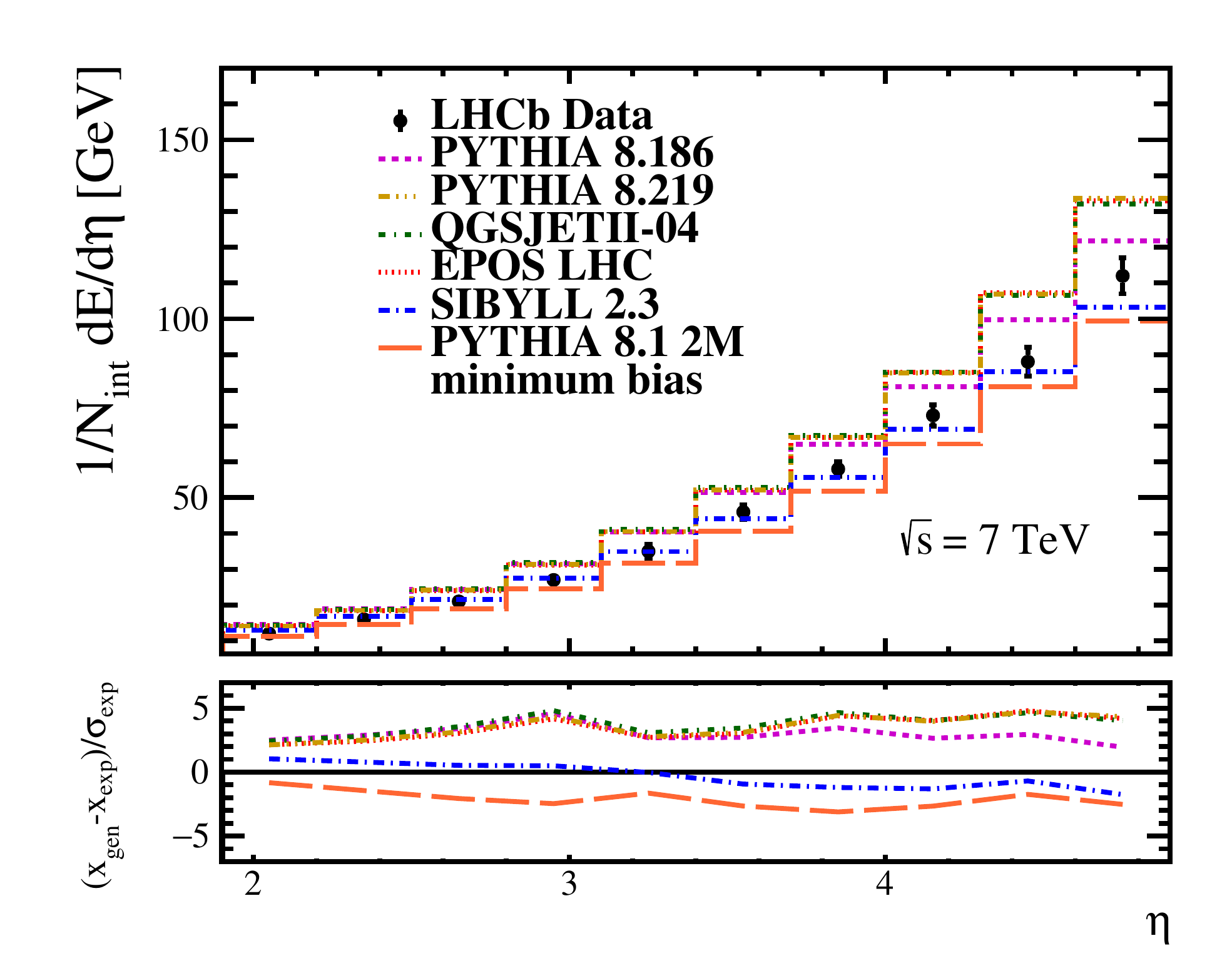} 
  \end{minipage}%%
  \begin{minipage}[b]{0.5\linewidth} 
    \centering
    \includegraphics[trim={0.8cm 0.cm 0.7cm 0.5cm},clip,width=0.9\textwidth]{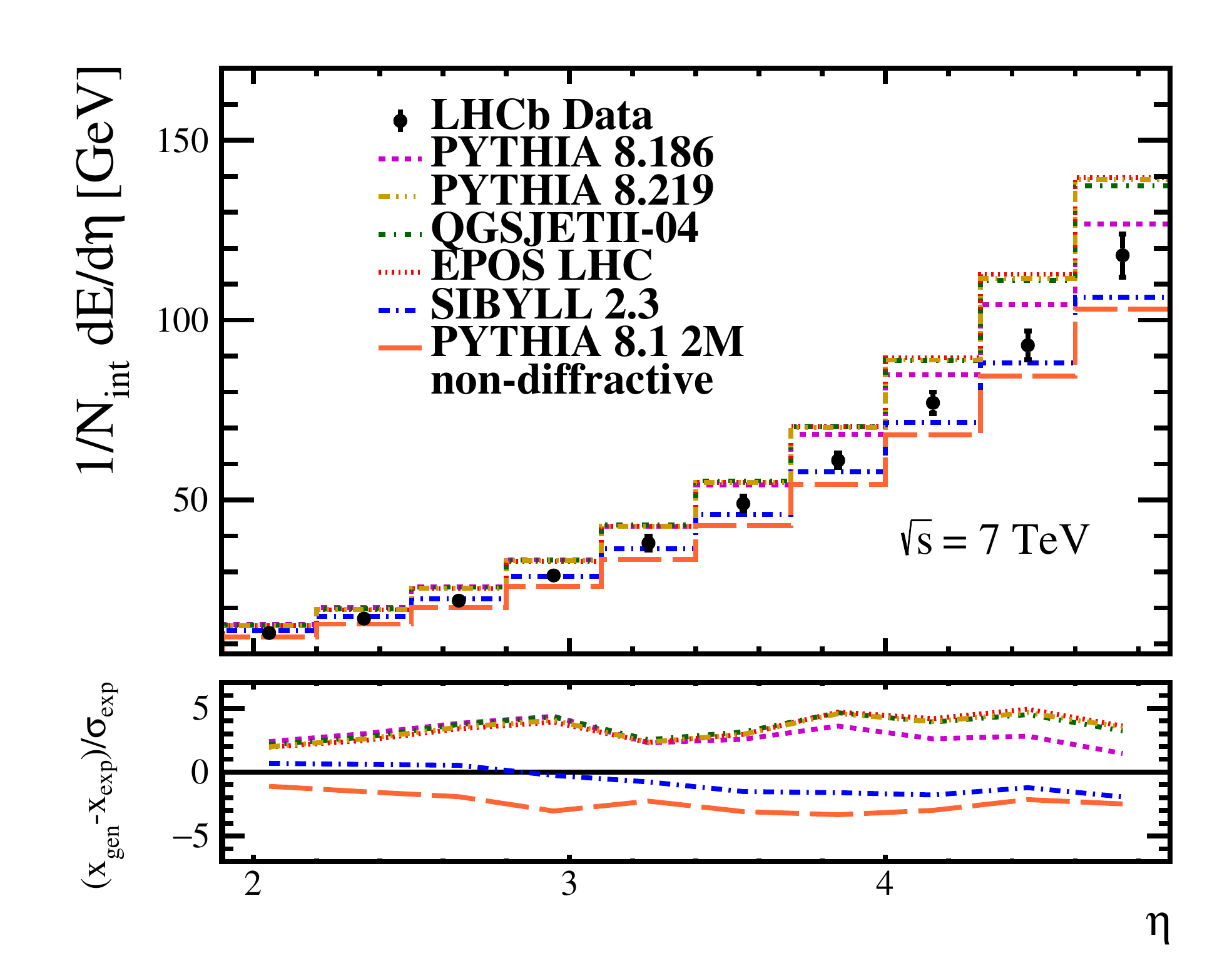} 
  \end{minipage}%%

  \begin{minipage}[b]{0.5\linewidth} 
    \centering
    \includegraphics[trim={0.8cm 0.cm 0.7cm 0.5cm},clip,width=0.9\textwidth]{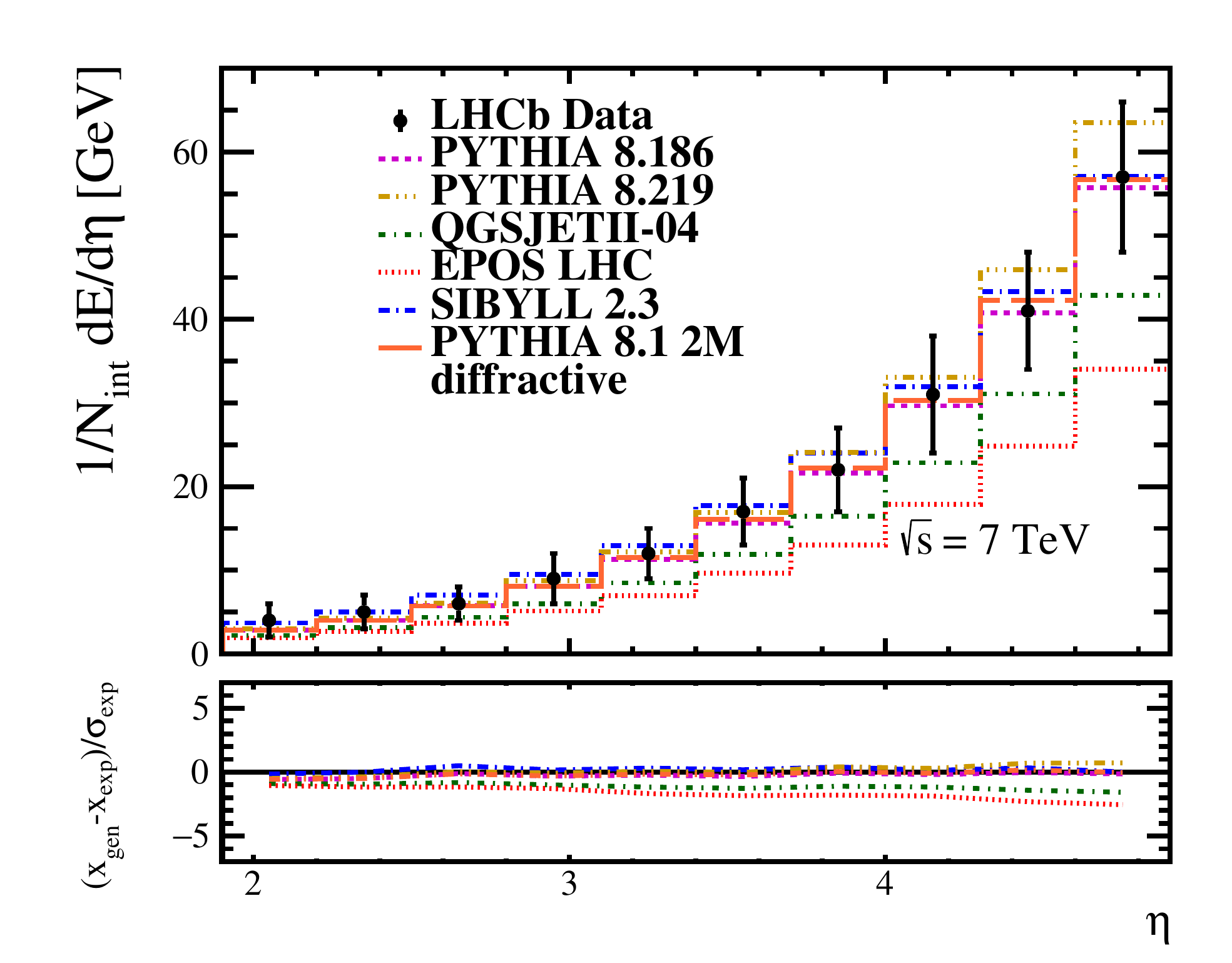} 
  \end{minipage}%%
  \begin{minipage}[b]{0.5\linewidth} 
    \centering
    \includegraphics[trim={0.8cm 0.cm 0.7cm 0.5cm},clip,width=0.9\textwidth]{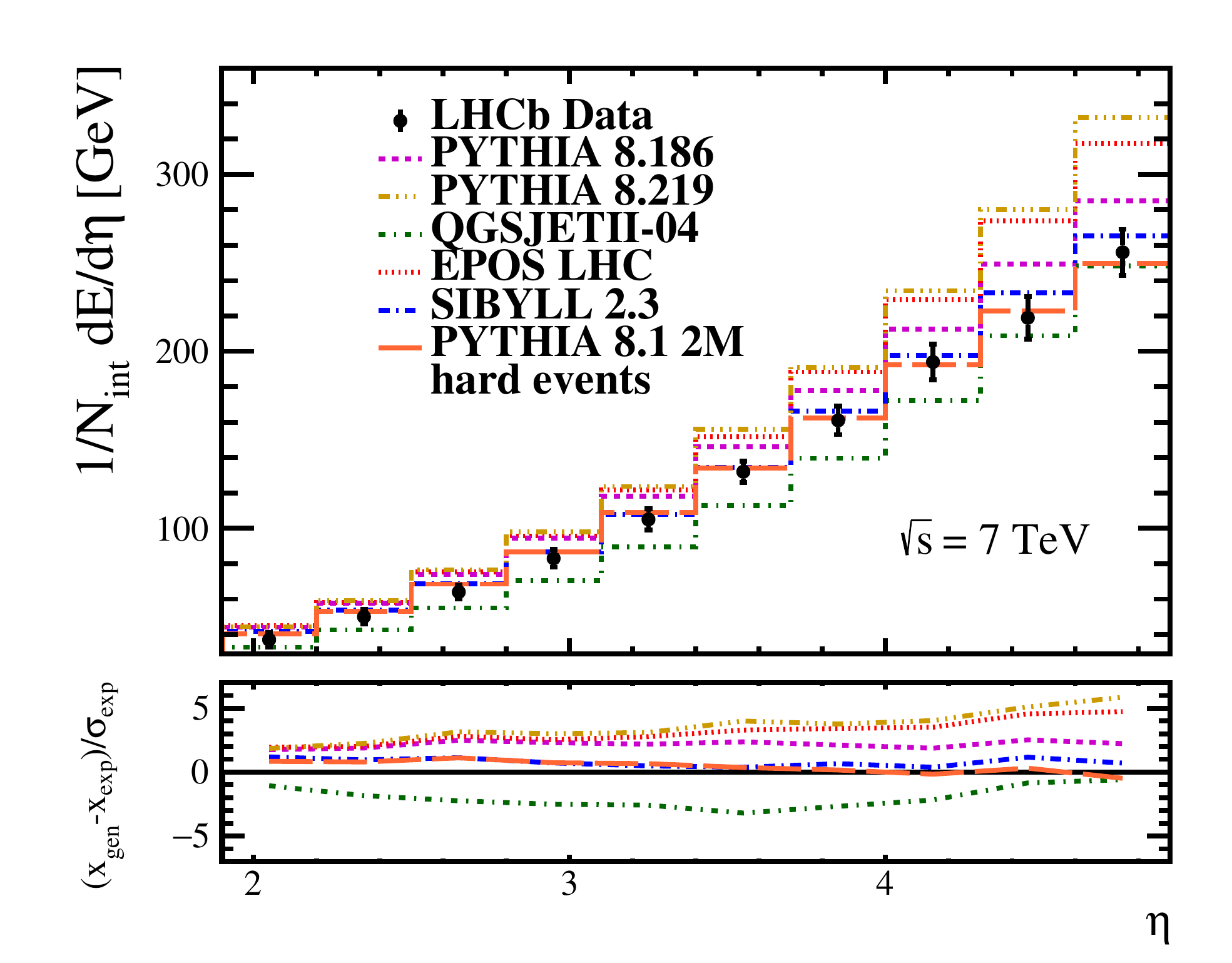} 
  \end{minipage}%%

  \caption{Charged energy flow for different classes of events from \emph{pp} collisions at $\sqrt{s}$ = 7 TeV. The LHCb data vertical bars represent the systematic uncertainty (the statistical uncertainty is negligible)  \cite{11}.}
  
\label{f1}

\end{figure}

The charged energy flow for different event classes is presented in Figure \ref{f1}. In Figures 1 and 2 from \cite{11} one can find the predictions for older pre-LHC tuned versions of the generators used in this study.

The predictions of PYTHIA 6's versions \cite{11} seem to be reasonably good in the central region (with the exception of diffractive events), but largely underestimate the measured values in the forward region in all cases. PYTHIA 8.135's predictions have a good description for the inclusive minimum bias, diffractive enriched and non-diffractive enriched event classes, but overestimate the measured values for the hard events.

PYTHIA 8.1 2M exhibits a slight decrease in overall values relative to version 8.135 (which uses the older Tune 1 \cite{10}) for the minimum bias, non-diffractive enriched and hard event classes. The description for the hard event class is improved, while for the other two event classes an underestimation trend is now observed. There is no major difference between the two versions for the diffractive event class.

With the exception of SIBYLL, a generator tuned to reproduce energy flow measurements, PYTHIA 8.186 seems to have the best description overall of the LHC-tuned generators. It's predictions for the diffractive enriched class are very similar to that of version 8.135, but for the rest of the event classes the predictions are further away from the measurements, exhibiting a constant overestimation trend.

PYTHIA 8.219 has a good description of the charged energy flow for the diffractive enriched class, being similar to that of version 8.186. One can see that the predictions tend to have an increased overestimation in the forward region, but are similar to the ones of version 8.186 in the central region. The differences can be explained by the 10\% increase in charged particle densities

\end{multicols}

%Energy flow...............................................................

%..............................................................

\begin{multicols}{2}

\noindent in the forward region implemented through the Monash 2013 tune \cite{19}.

EPOS 1.99's predictions \cite{11} describe reasonably well the charged energy flow for inclusive minimum bias, hard and non-diffractive enriched event classes, slightly overestimating the measurements in the last two bins, and it underestimates the charged energy flow for diffractive processes in the forward region.

EPOS LHC's predictions are very similar to the ones of PYTHIA 8.219 for all event classes except the diffractive enriched class, where, similarly to the previous version, it underestimates the charged energy flow. As one can see in the restricted minimum bias plot, the apparent overestimation of the soft process component is similar to the one of PYTHIA 8.219. Compared to the previous version, we observe a worsening of the predictions (except for diffractive events). EPOS LHC shows an overall overestimation of the measurements with an increasing trend towards the forward pseudorapidity region.

The predictions of QGSJET01 and QGSJETII-03 from \cite{11} are similar for the inclusive minimum bias class and they overestimate the charged energy flow. QGJSET01 has a better description of the diffractive and hard events class in the central region, but tends to overestimate the measurements for the hard events and underestimate them for the diffractive ones in the forward region. The general trend of QGSJETII-03 is of underestimating for the hard events.

The prediction of QGSJETII-04 is similar to that of the previous versions for the inclusive minimum bias event class. The description of the charged energy flow for hard events is more underestimated than in the case of QGSJETII-03. The diffractive component's description is similar to that of QGSJETII-03, but with a slightly larger underestimation trend. For the rest of the event classes the differences with respect to the measured LHCb charged energy flow are significant. Although the absolute values are rather clearly far from the experimental values, the shapes are well described. QGSJETII-04 is very similar to EPOS LHC and PYTHIA 8.219 in it's description of the charged energy flow for inclusive minimum bias and non-diffractive enriched event classes.

SIBYLL 2.1's prediction \cite{11} describes very well the measurements for inclusive minimum bias events. It also has a reasonably good description for the diffractive events, the values being within the error bars, although an underestimation trend can be seen. The hard events component is well described in the central region, but it is overestimated in the forward region.

SIBYLL 2.3 seems to have the best prediction for all event classes (on par with PYTHIA 8.186 for the diffractive enriched class). It can be seen that it has a slight underestimation trend in the forward region in the case of inclusive minimum bias and non-diffractive enriched event classes.

As one can see in Table \ref{t1}, PYTHIA 8.219 and EPOS have similar ratios of hard events, but the number of visible events and the ratio of diffractive events are smaller for EPOS. PYTHIA 8.186's ratio of hard events is larger than version 8.219's one, but the ratios of diffractive events are close indicating that the mechanisms of diffractive processes are similar. QGSJETII-04's ratio of hard events is sensibly larger than the rest and the ratio of diffractive events is smaller, so the hard process component seems to be larger for this generator. Likewise, SIBYLL's hard process component is larger than PYTHIA's and EPOS's one.

\begin{figure}[t!] 

\center
  \begin{minipage}[b]{0.5\linewidth} 
    \centering
    \includegraphics[trim={0.8cm 0.cm 0.9cm 0.5cm},clip,width=0.9\textwidth]{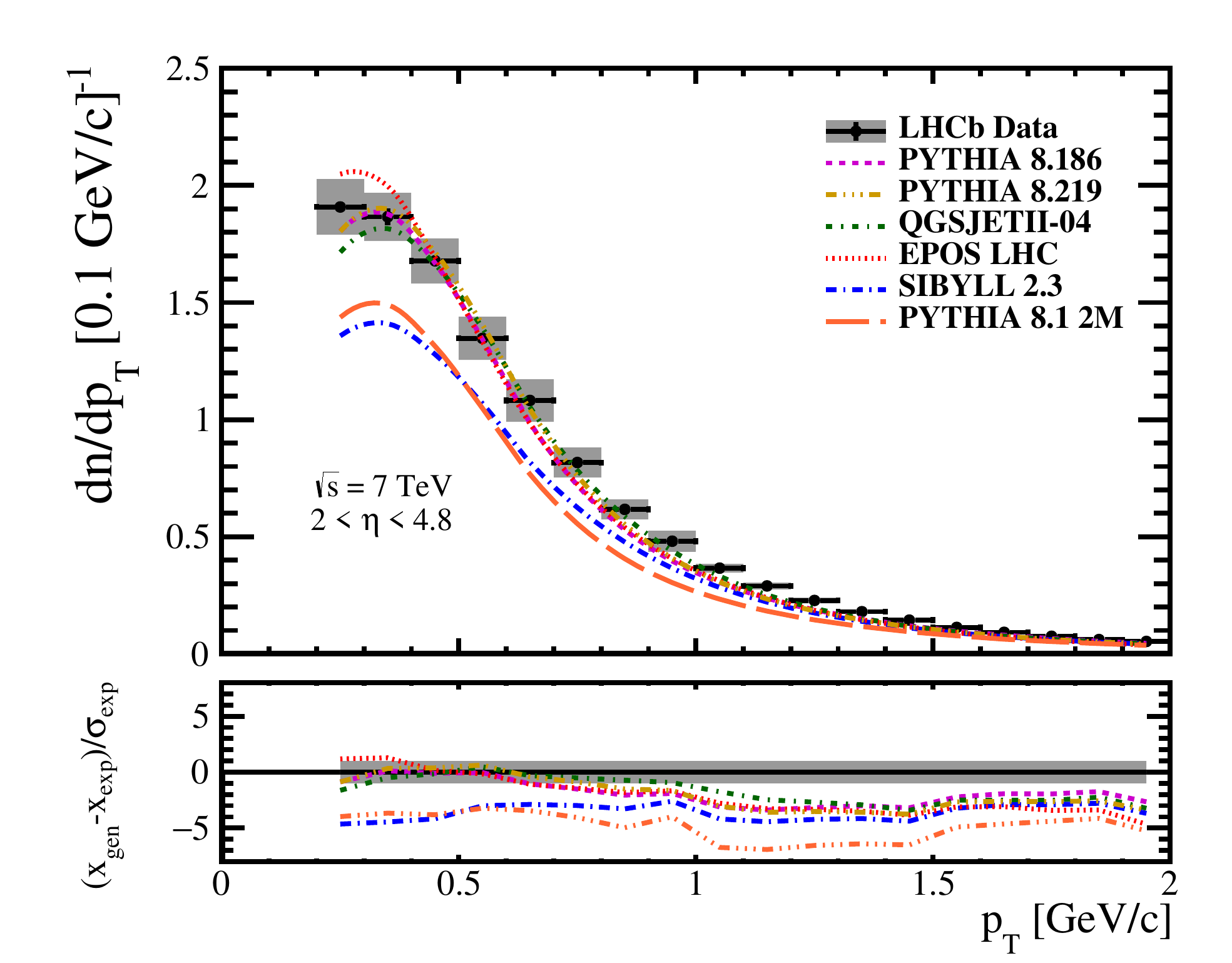} 
  \end{minipage}%%
  
  \begin{minipage}[b]{0.5\linewidth} 
    \centering
    \includegraphics[trim={0.8cm 0.cm 0.9cm 0.5cm},clip,width=0.9\textwidth]{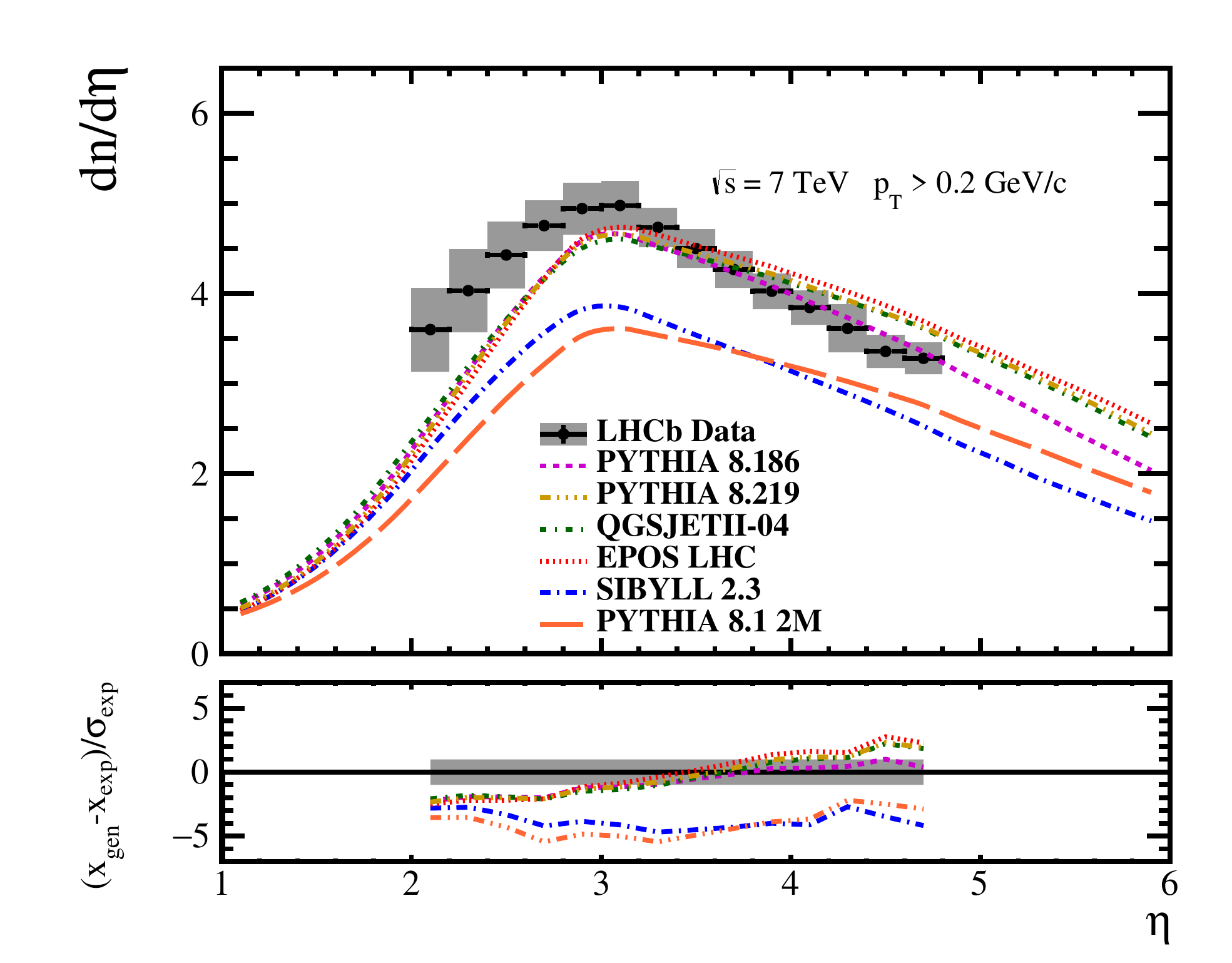} 
  \end{minipage}%%
   \begin{minipage}[b]{0.5\linewidth} 
    \centering
    \includegraphics[trim={0.8cm 0.cm 0.9cm 0.5cm},clip,width=0.9\textwidth]{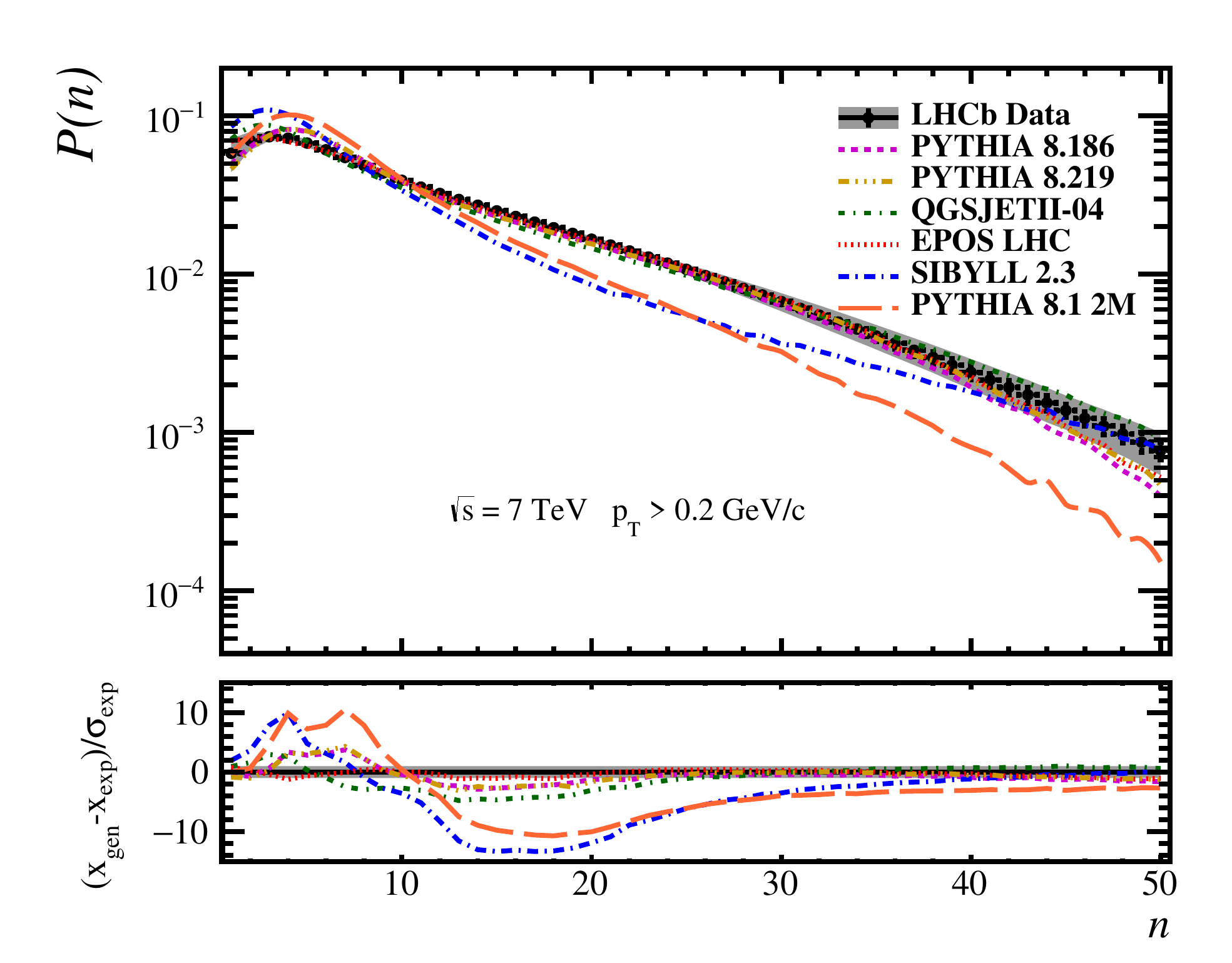} 
  \end{minipage}%%

  \caption{Transverse momentum, pseudorapidity and multiplicity distributions for prompt charged particles in the kinematic region of $2 <\eta < 4.8$, $p\ge2$ GeV/c and $p_{T}>0.2$ GeV/c at $\sqrt{s}$ = 7 TeV. The vertical bars represent the statistical error and the grey bands represent the combined uncertainties (statistical and systematic) \cite{14}.}

\label{f3}

\end{figure}

As one can see in the transverse momentum plot from Figure \ref{f3}, PYTHIA and QGSJET predictions are similar in shape. There is no major difference between PYTHIA's LHC-tuned versions. PYTHIA and EPOS predictions are rather similar in the interval 0.5-1.5 GeV/c. QGSJET's prediction seems closest to the LHCb measurements, but for all generators there are visible differences in absolute scale, especially in the hard part of the spectrum. SIBYLL-generated spectrum has a shape which approaches the experimental one, but the absolute values differ significantly. The shapes of the spectrums generated with QGSJET, EPOS and both versions of PYTHIA are close to the experimental one.

In the pseudorapidity plot from Figure \ref{f3} one can see that all the predictions cluster together at low values as the models were tuned using measurements from central LHC experiments. QGSJET, EPOS and PYTHIA 8.2 underestimate the measurements for values below $\eta=$ 3.5 and overestimates them in the forward region (where they also remain clustered together). PYTHIA 8.1 also underestimates the measurements in the central region, but the prediction in the forward region seems to be reasonably good. SIBYLL largely underestimates the measurements across the whole range.

For the (probability density of) multiplicity distribution from Figure \ref{f3}, the closest prediction seems to be the one of EPOS. All LHC-tuned generators reproduce the measurements well for this distribution, except SIBYLL which deviates significantly. One can see that EPOS's prediction clusters together with PYTHIA estimates in the medium-high multiplicity region. For values below $n_{ch}$ = 10 EPOS seems to be better than PYTHIA. QGSJET's prediction is close to the ones of EPOS and PYTHIA, but the underestimation at low multiplicities in the interval $n_{ch}$ = 10-20 is larger, the deviations from the measurements ranging between $\sim 3-5$ $\sigma$. SIBYLL's prediction very strongly favours low multiplicities, but gets closer to the measured values towards high multiplicities.

\vspace{0.5 cm}

\begin{center}

\tabcaption{ \label{t1}  Number of visible events for different event classes. $N_{MB}$, the number of visible minimum bias events, is expressed as percentages from the total number of generated inelastic events $N_{gen}=10^{6}$. $N_{hard}$ and $N_{dif}$, the numbers of visible hard and diffractive events, respectively, are expressed as percentages of $N_{MB}$.}

\begin{tabular}{l c c c} \toprule
    {Generator} & {$N_{MB}$} & {$N_{hard}$} & {$N_{dif}$} \\ \midrule

    PYTHIA 8.186  & 88.20 \% & 5.63 \% & 7.04 \% \\
   
    PYTHIA 8.219  & 88.11 \% & 5.05 \% & 7.10 \% \\
    
    EPOS LHC      & 84.92 \% & 4.87 \% & 6.26 \% \\
    
	QGSJETII-04   & 86.72 \% & 7.94 \% & 5.52 \% \\

    SIBYLL 2.3    & 89.55 \% & 6.43 \% & 6.47 \%  \\ 
    
    PYTHIA 8.1 2M    & 86.89 \% & 5.08 \% & 7.97 \%  \\ \bottomrule

\end{tabular}

\end{center}

\vspace{0.5 cm}

\begin{figure}[H] 

  \begin{minipage}[b]{0.5\linewidth} 
    \centering
    \includegraphics[trim={0.8cm 0.cm 0.7cm 0.5cm},clip,width=0.9\textwidth]{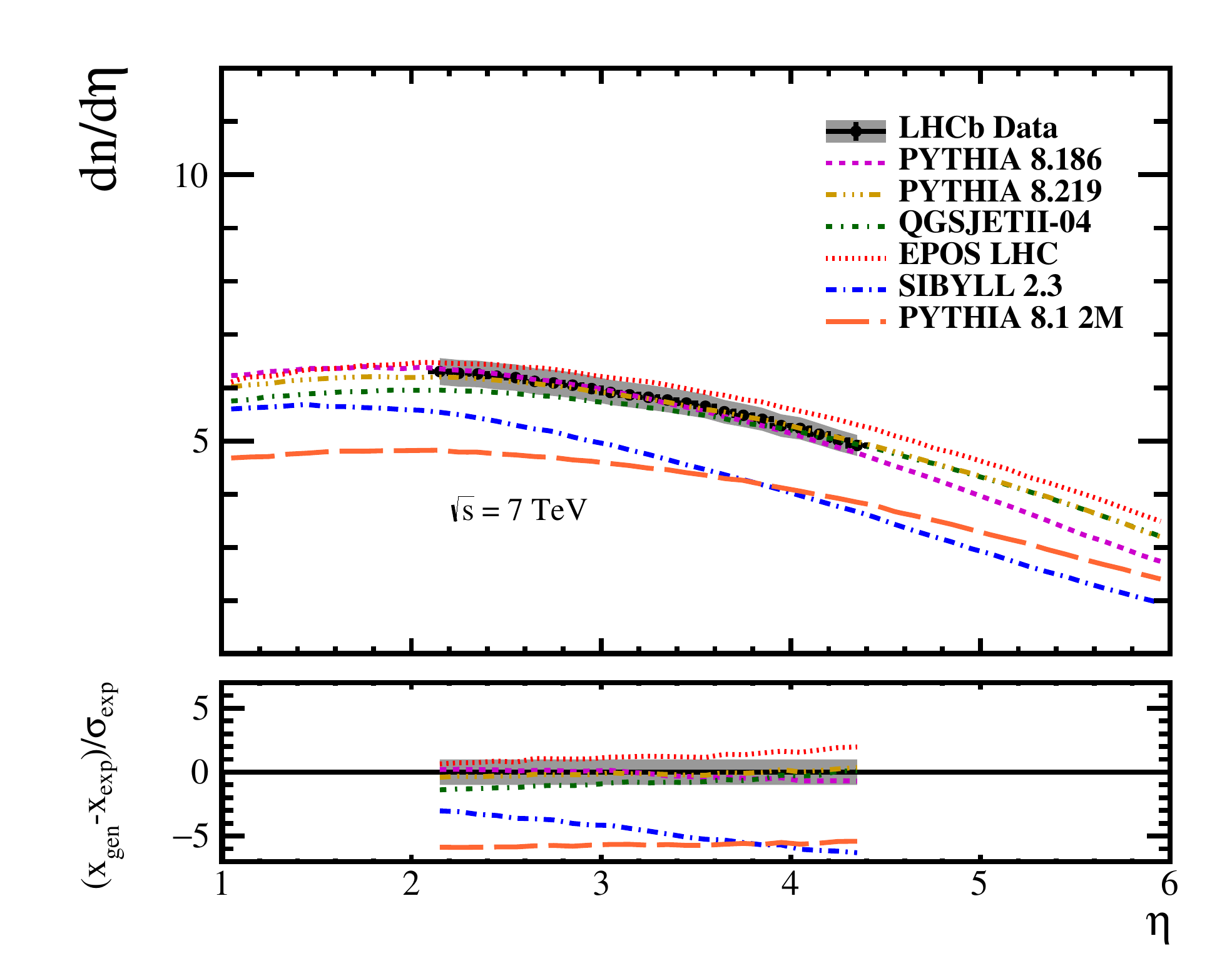} 
  \end{minipage}%%
  \begin{minipage}[b]{0.5\linewidth} 
    \centering
    \includegraphics[trim={0.8cm 0.cm 0.7cm 0.5cm},clip,width=0.9\textwidth]{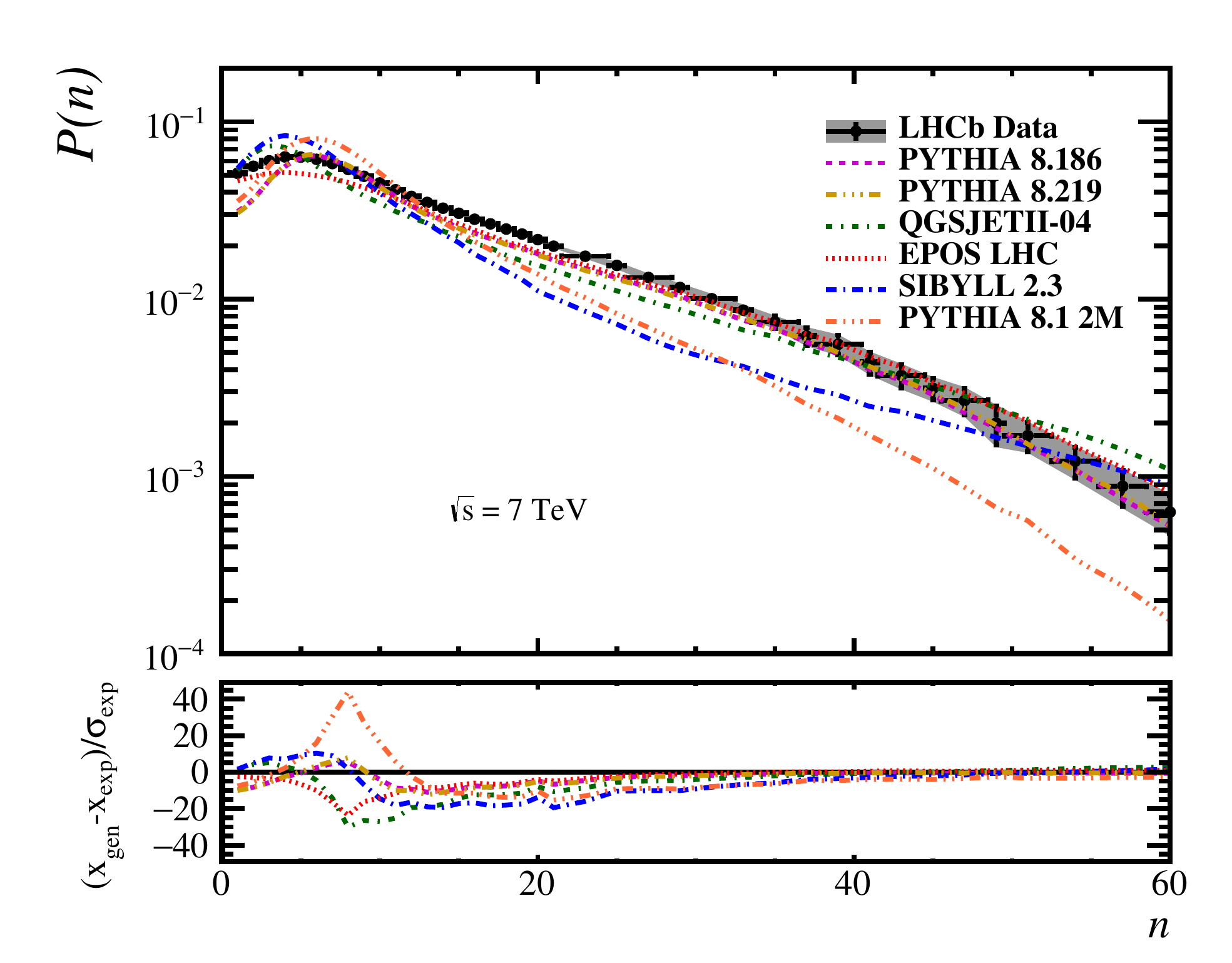} 
  \end{minipage}%%

  \caption{Pseudorapidity and multiplicity distributions for prompt charged particles in the kinematic region of $2 <\eta < 4.5$ at $\sqrt{s}$ = 7 TeV. The vertical bars represent the statistical error and the grey bands represent the combined uncertainties (statistical and systematic) \cite{25}.}

\label{f4}

\end{figure}

The pseudorapidity distribution from Figure \ref{f4} is best described by PYTHIA 8.186. PYTHIA 8.219's prediction is close, too. EPOS and QGSJET estimates are a bit further away from the experimental values. SIBYLL's prediction is significantly different both in absolute value as well as shape of the distribution. With the exception of SIBYLL, the clustering of the predictions can be seen in the central pseudorapidity region, indicating the tuning was done using similar measurements. The prediction of EPOS describes the measurements reasonably well in the central region ($2<\eta<2.5$), but it diverges upwards from the measured values in the forward region. This effect of overestimation in the forward region is similar to the one seen in Figure \ref{f3}. QGSJET slightly underestimates the measurements in the central region, but gets closer in the forward region (overlapping with PYTHIA 8.219).

The multiplicity distribution is not perfectly described by any of the generators, but one can see that the predictions of EPOS and PYTHIA seem to get better at higher multiplicities, as we have also seen for the previous multiplicity distribution. The distributions generated with SIBYLL and QGSJET are significantly different from the experimental ones.

\end{multicols}

%Event..............................................................

%Minbias...................................................

%Hard...................................................

\begin{figure}[t!]

   \begin{minipage}[b]{0.5\linewidth} 
    \centering
    \includegraphics[trim={0.8cm 0.cm 0.7cm 0.5cm},clip,width=0.9\textwidth]{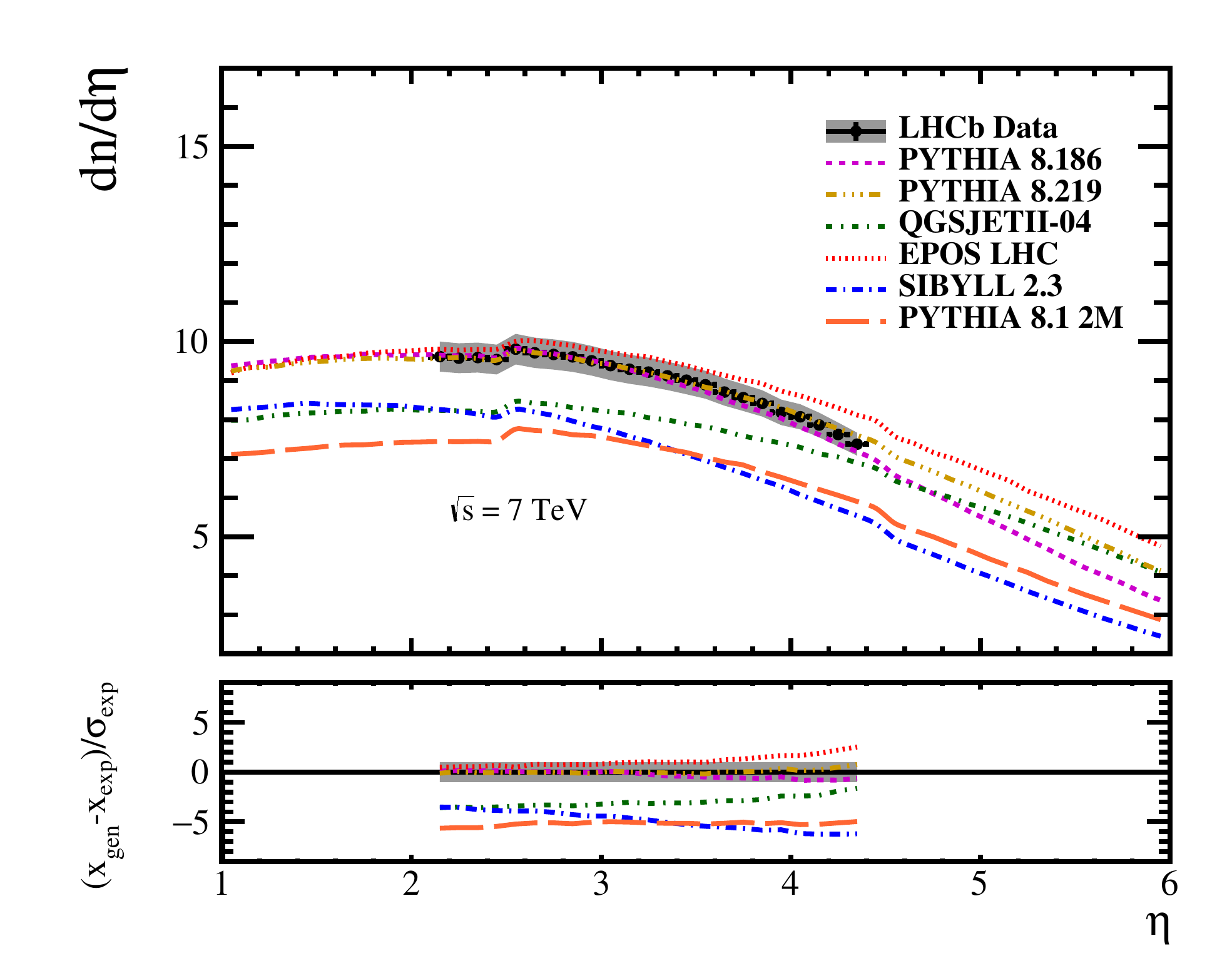} 
  \end{minipage}%%
   \begin{minipage}[b]{0.5\linewidth} 
    \centering
    \includegraphics[trim={0.8cm 0.cm 0.7cm 0.5cm},clip,width=0.9\textwidth]{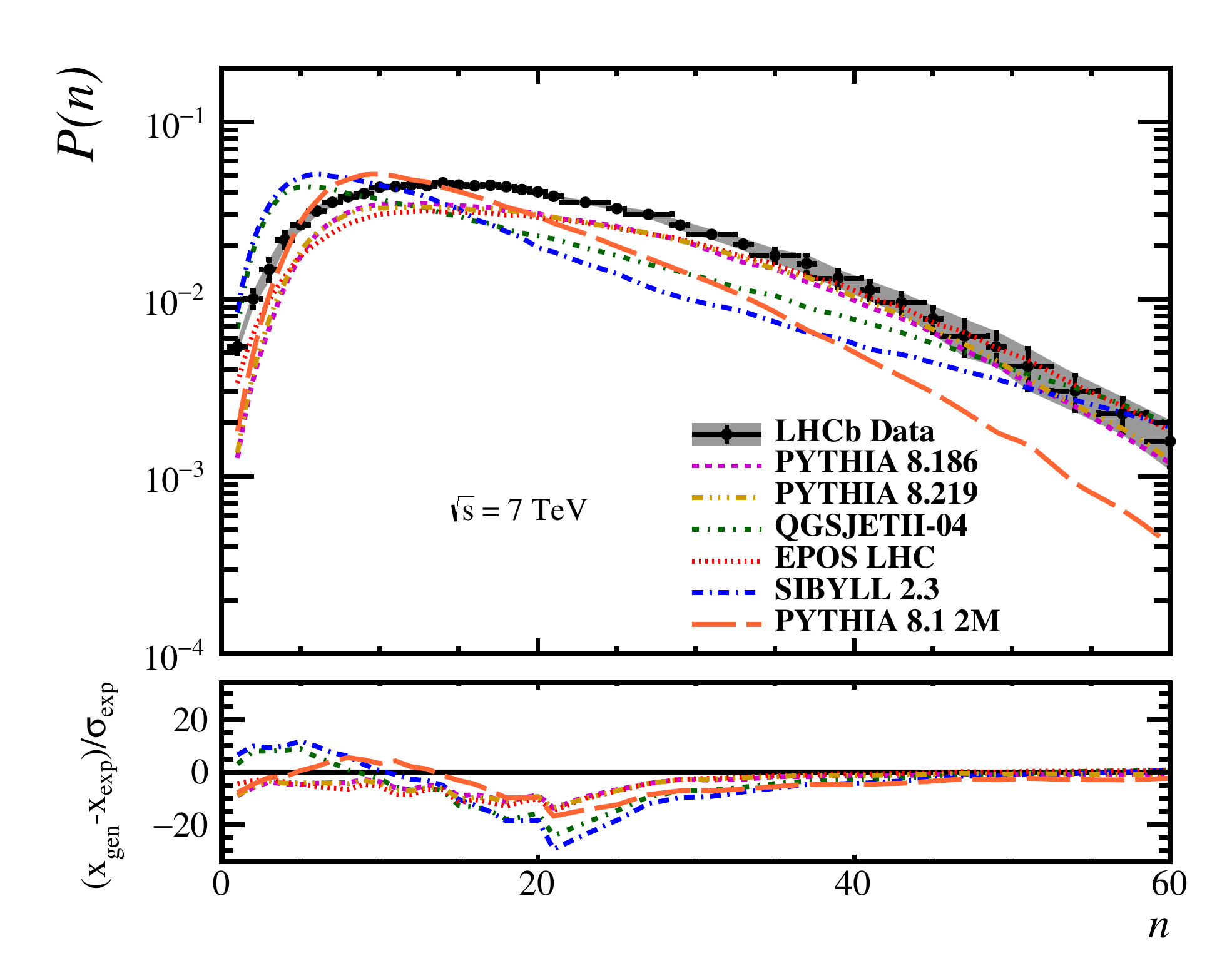} 
  \end{minipage}%%

  \caption{Pseudorapidity and multiplicity distributions for prompt charged particles in the kinematic region of $2 <\eta < 4.5$ from ``hard" events at $\sqrt{s}$ = 7 TeV. The vertical bars represent the statistical error and the grey bands represent the combined uncertainties (statistical and systematic) \cite{25}.}

\label{f5}

\end{figure}

\begin{multicols}{2}

%Totem....................................

\begin{center} 

    \includegraphics[trim={0.8cm 0.cm 0.7cm 0.5cm},clip,width=0.4\textwidth]{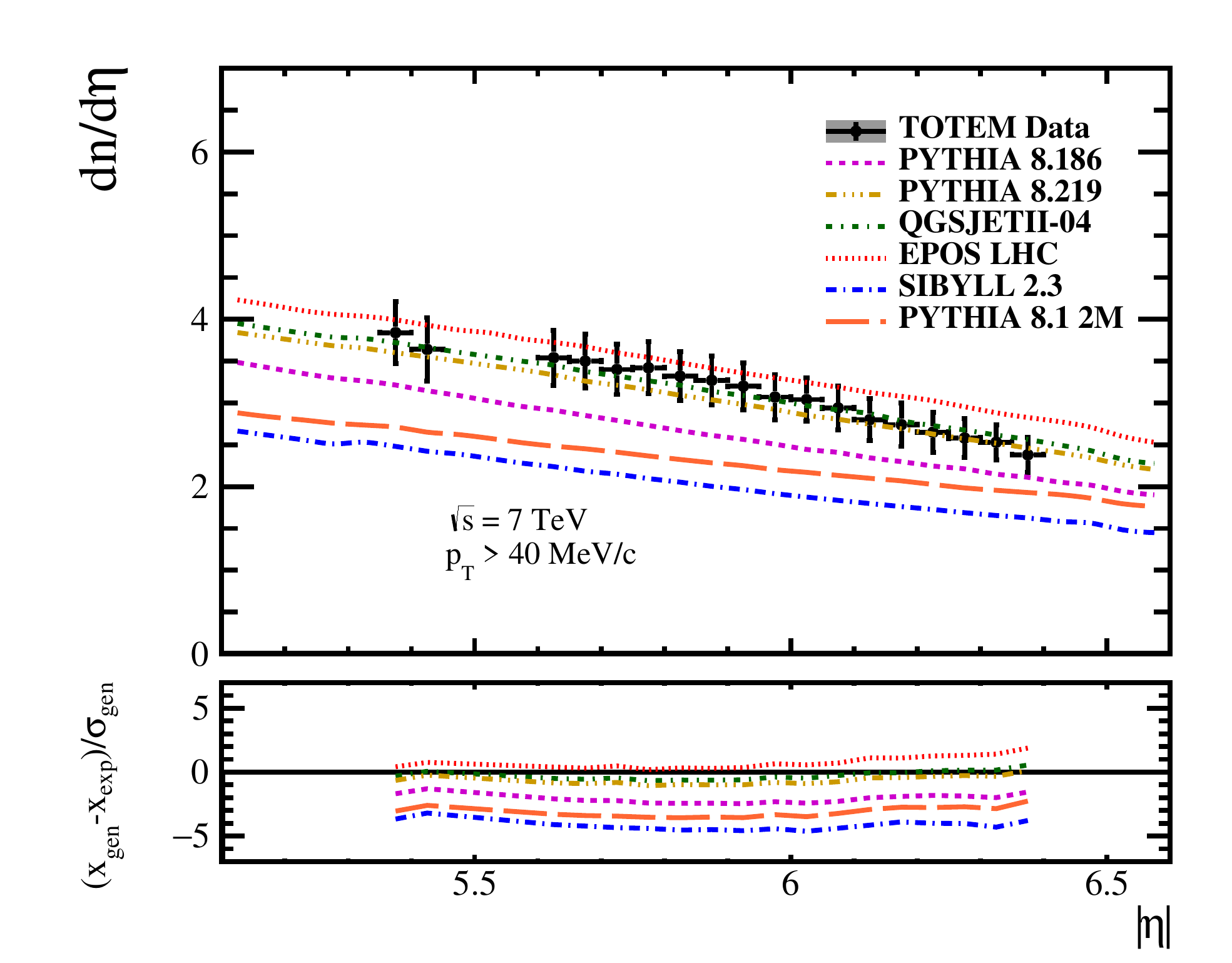} 
\figcaption{\label{f6}   Prompt charged-particle pseudorapidity distribution in the kinematic region of $p_{T}>$ 40 MeV/c and 5.3 $\le|\eta|\le$ 6.5 at $\sqrt{s}$ = 7 TeV. The error bars represent the combined statistical and systematic errors \cite{27}.}
\end{center}

The pseudorapidity plot from Figure \ref{f5} shows a good agreement between PYTHIA versions and the LHCb measurements. EPOS also has a good description of the measurements in the central region, but diverges upwards in the forward region. SIBYLL's prediction is similar to the one of QGSJET at low rapidity, but they diverge in the forward region and are both far from the experimental distribution. The discontinuity at $\eta=2.5$ is due to the hard event selection criterion of a minimum of one particle with $2.5\leq \eta \leq 4.5$ and $p_{T}\geq1$ GeV/c \cite{25}.

As in Figure \ref{f4}, the multiplicity distribution is not well described by the generators with PYTHIA and EPOS being closest to the measurements.

As can be seen in Figure \ref{f6} the best predictions are the ones of QGSJET, EPOS and PYTHIA 8.219. All the generated shapes and spectrum slope agree well with the ones of the experimental distribution.

In the pseudorapidity plots from figures \ref{f4}-\ref{f6} it can be seen that the predictions of PYTHIA 8.1 2M largely underestimate the measurements. The differences between the predictions of PYTHIA with Tune 2M and the two LHC tunes are large in the central region and exhibit a converging trend towards higher pseudorapidity. The multiplicity plots from figures \ref{f3}-\ref{f6} are rather clearly not well reproduced by PYTHIA 8.1 2M's prediction which favours very low multiplicities.

\vspace{0.5 cm}

\begin{center}

\tabcaption{ \label{t2}  Number of events with a minimum of $n_{ch}\geq1$ in $2<\eta<4.5$ expressed as percentages from the total number of generated inelastic events $N_{gen}=10^{6}$. Hard events require a minimum of one charged particle with $p_{T}\geq1$ GeV/c in $2.5<\eta<4.5$.}

\begin{tabular}{l c c} \toprule
    {Generator} & minimum bias & hard events \\ 
     &  & [\% of minbias] \\ \midrule

    PYTHIA 8.186  & 87.28 \% & 43.90 \%  \\
   
    PYTHIA 8.219  & 87.17 \% & 42.83 \%  \\
    
    EPOS LHC      & 83.81 \% & 44.86 \%  \\
    
	QGSJETII-04   & 85.57 \% & 54.01 \%  \\

    SIBYLL 2.3    & 88.19 \% & 46.68 \%   \\ 

    PYTHIA 8.1 2M    & 85.87 \% & 37.37 \%   \\ \bottomrule

\end{tabular}
\end{center}

\vspace{0.5 cm}

The ratios of hard events for PYTHIA and EPOS, given in Table \ref{t2}, are close, suggesting a similarity between the descriptions of hard processes. SIBYLL's ratio is slightly higher than the previous generators. QGSJET's ratio of hard events is considerably higher than the ratios of the other generators, so again one can see that it favours the hard processes.

\begin{figure}[t!]

  \begin{minipage}[b]{0.5\linewidth} 
    \centering
    \includegraphics[trim={0.9cm 0.8cm 0.9cm 1cm},clip,width=0.73\textwidth]{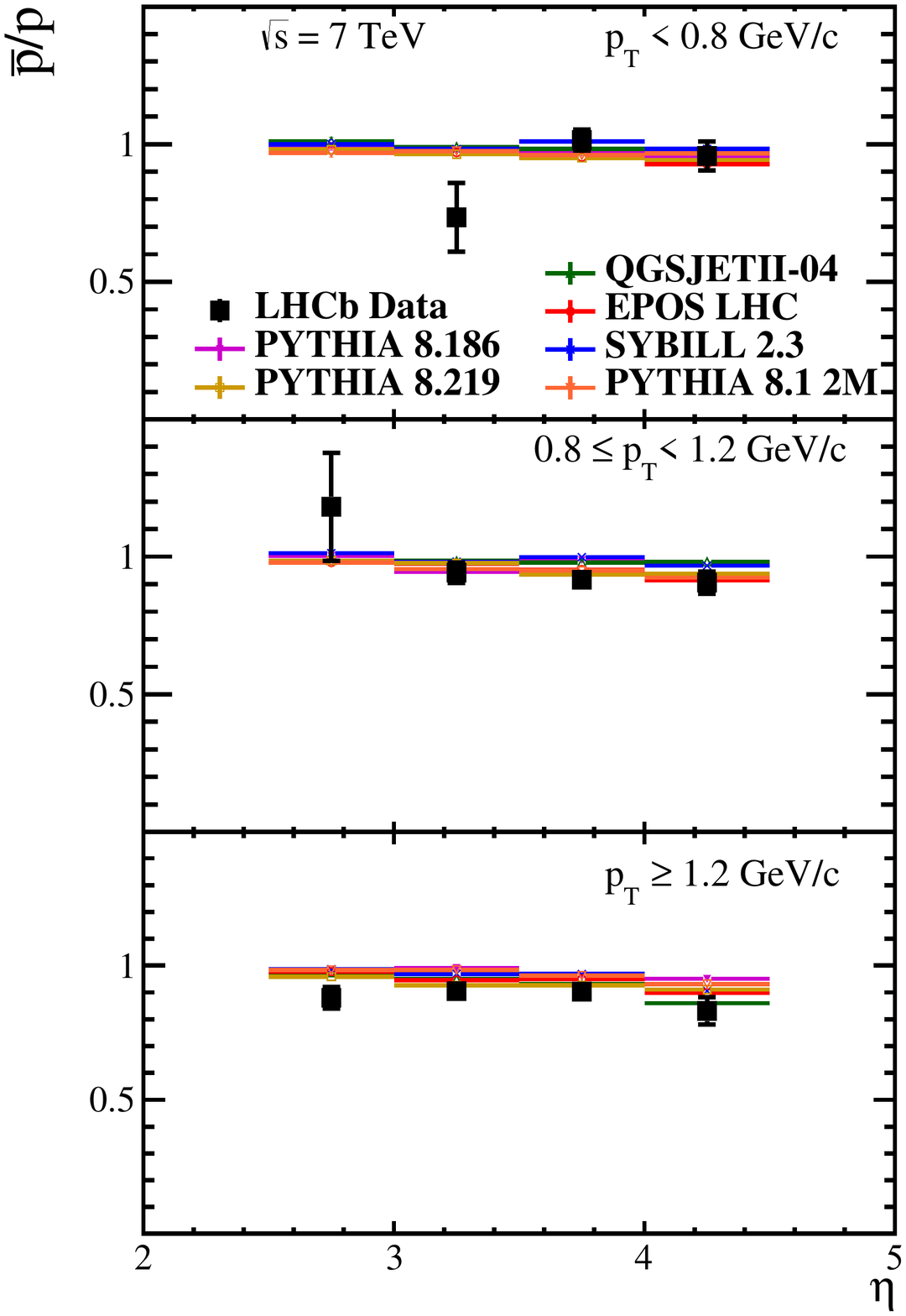} 
  \end{minipage}%%
  \begin{minipage}[b]{0.5\linewidth} 
    \centering
    \includegraphics[trim={0.9cm 0.8cm 0.9cm 1cm},clip,width=0.73\textwidth]{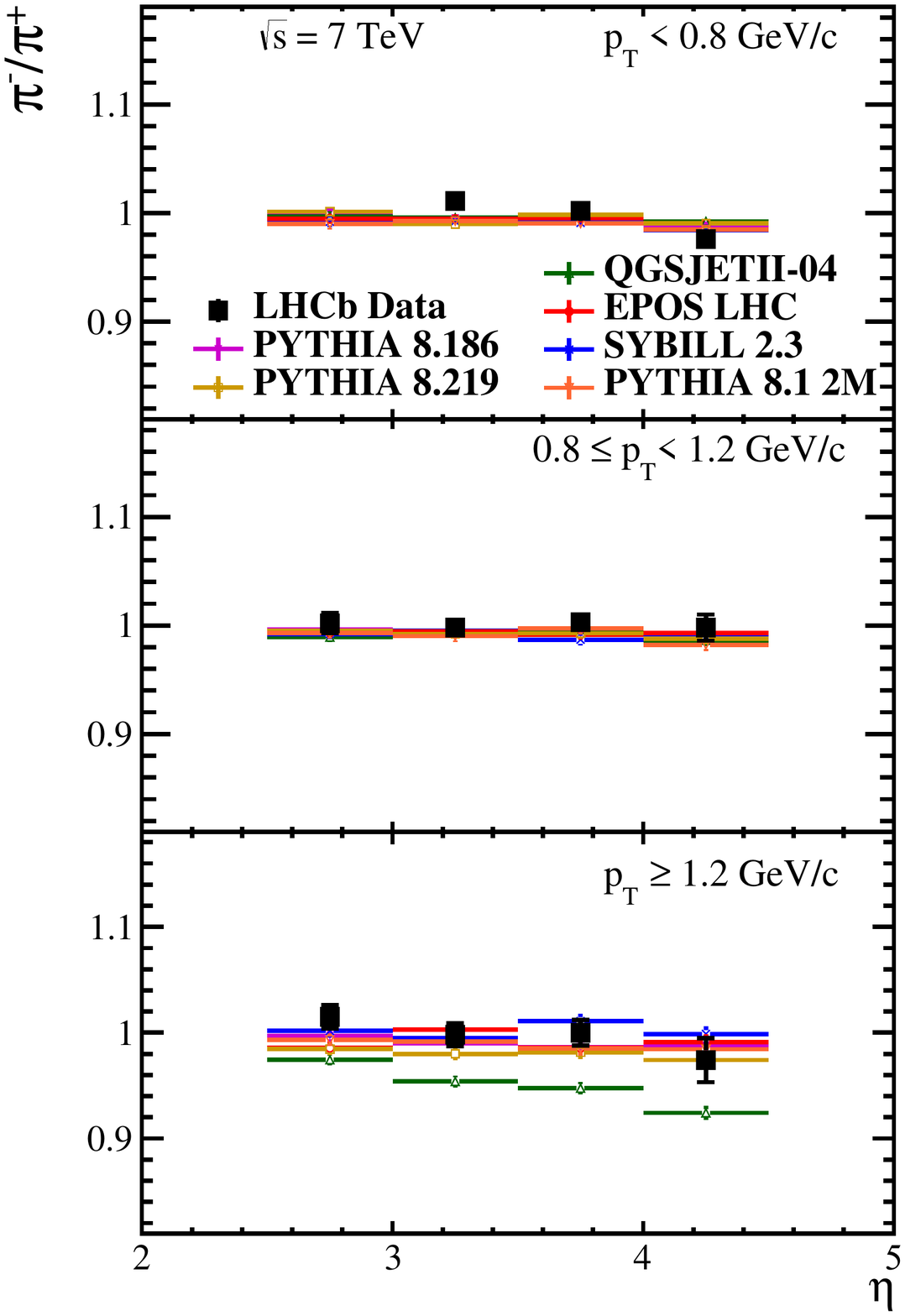} 
  \end{minipage}%%

  \caption{Prompt charged-hadron ratios as a function of pseudorapidity in the kinematic region of $2.5\le\eta\le4.5$ and $p\ge5$ GeV/c in various $p_{T}$  intervals at $\sqrt{s}$ = 7 TeV. The LHCb data vertical bars represent the combined statistical and systematic uncertainties \cite{12}.}

\label{f7}

\end{figure}

\begin{figure}[h]

  \begin{minipage}[b]{0.5\linewidth} 
    \centering
    \includegraphics[trim={0.9cm 0.8cm 0.9cm 1cm},clip,width=0.73\textwidth]{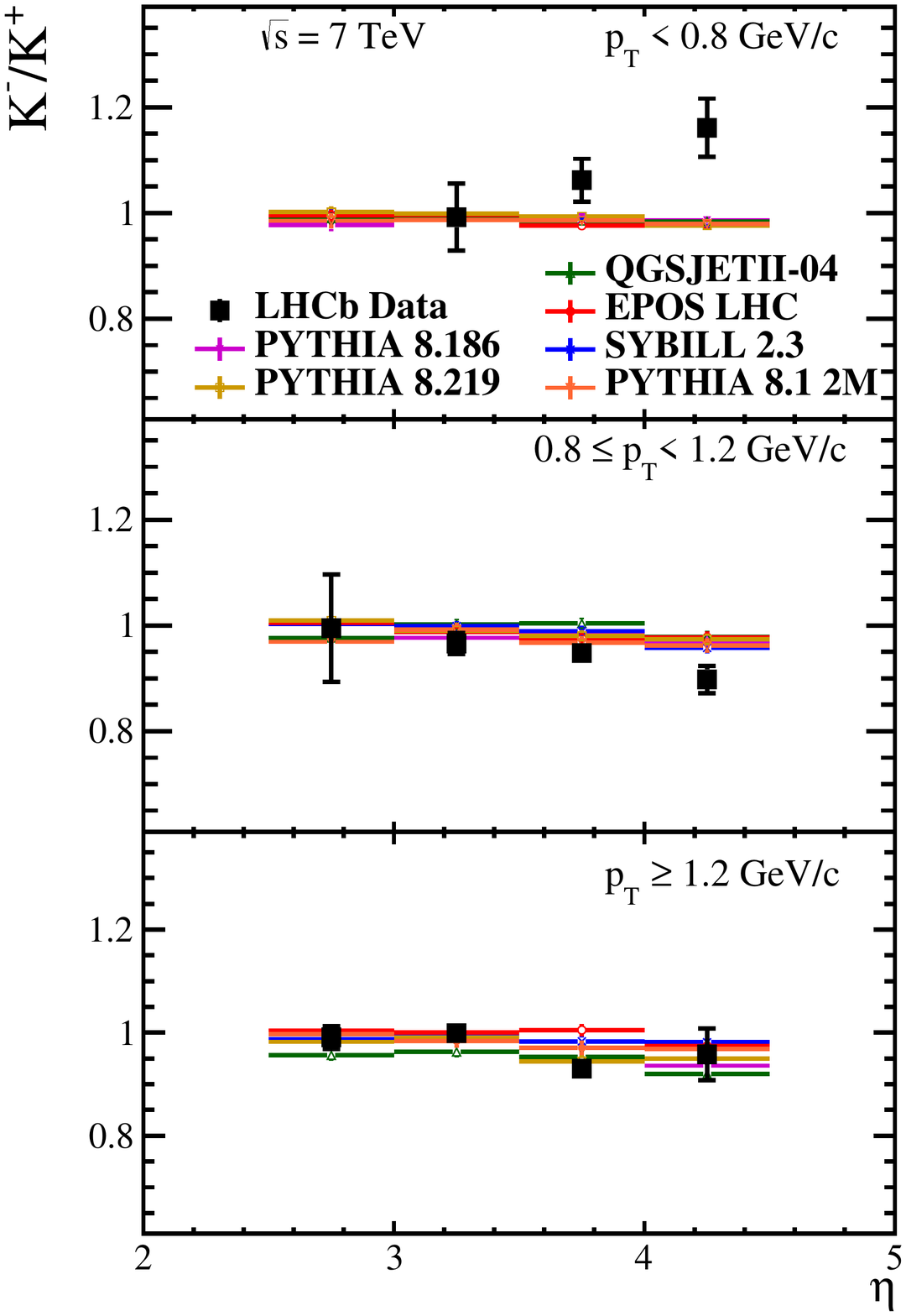} 
  \end{minipage}%%
  \begin{minipage}[b]{0.5\linewidth} 
    \centering
    \includegraphics[trim={0.9cm 0.8cm 0.9cm 1cm},clip,width=0.73\textwidth]{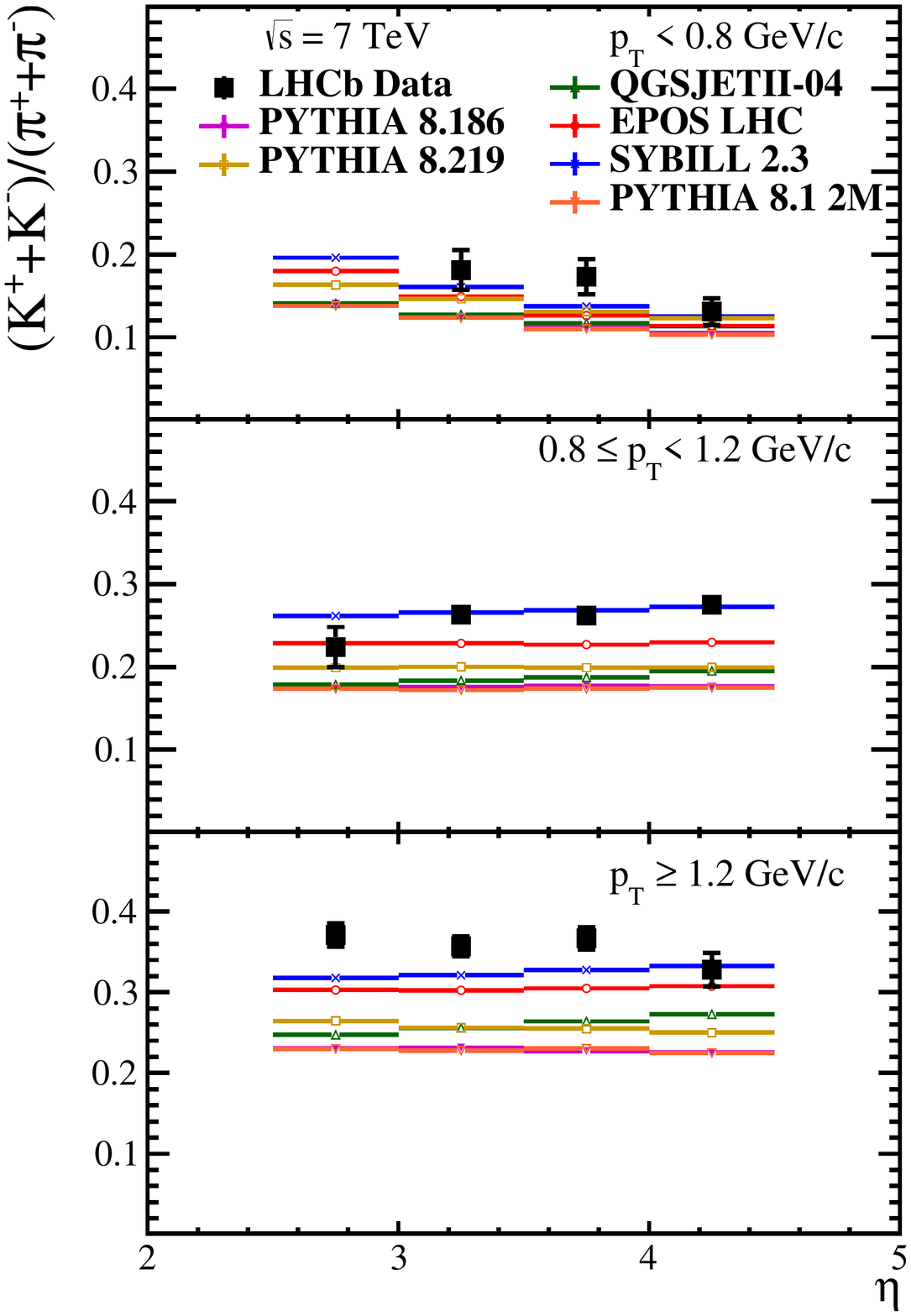} 
  \end{minipage}%%

  \caption{Prompt charged-hadron ratios as a function of pseudorapidity in the kinematic region of $2.5\le\eta\le4.5$ and $p\ge5$ GeV/c in various $p_{T}$  intervals at $\sqrt{s}$ = 7 TeV. The LHCb data vertical bars represent the combined statistical and systematic uncertainties \cite{12}.}

\label{f8}

\end{figure}

The plot for the $\bar{p}/p$ ratio is shown in Figure \ref{f7}. All predictions have the same trend of apparent decrease towards the beamline and it can be said that the ratio is reasonably well described. The $\pi^{-}/\pi^{+}$ ratio which is shown in the same figure is also well described by all generators with the exception of QGSJET for the high $p_{T}$ region, where it seems to show a charge asymmetry between $\pi^{+}$ and $\pi^{-}$. Also, all the predictions seem to cluster together, again with the exception of QGSJET at high $p_{T}$. The $K^{-}/K^{+}$ ratio shown in Figure \ref{f8} is fairly well described by all generators.

The closest prediction for the $(K^{+}+K^{-})/(\pi^{+}+\pi^{-})$ (shown in the same figure) seems to be that of SIBYLL followed by the one of EPOS, yet, overall all generators fail to describe this measurement. In the high $p_{T}$ range, QGSJET underestimates the measurements and has a pronounced ascending trend.

\begin{figure}[t!]

  \begin{minipage}[b]{0.5\linewidth} 
    \centering
    \includegraphics[trim={0.9cm 0.8cm 0.9cm 1cm},clip,width=0.73\textwidth]{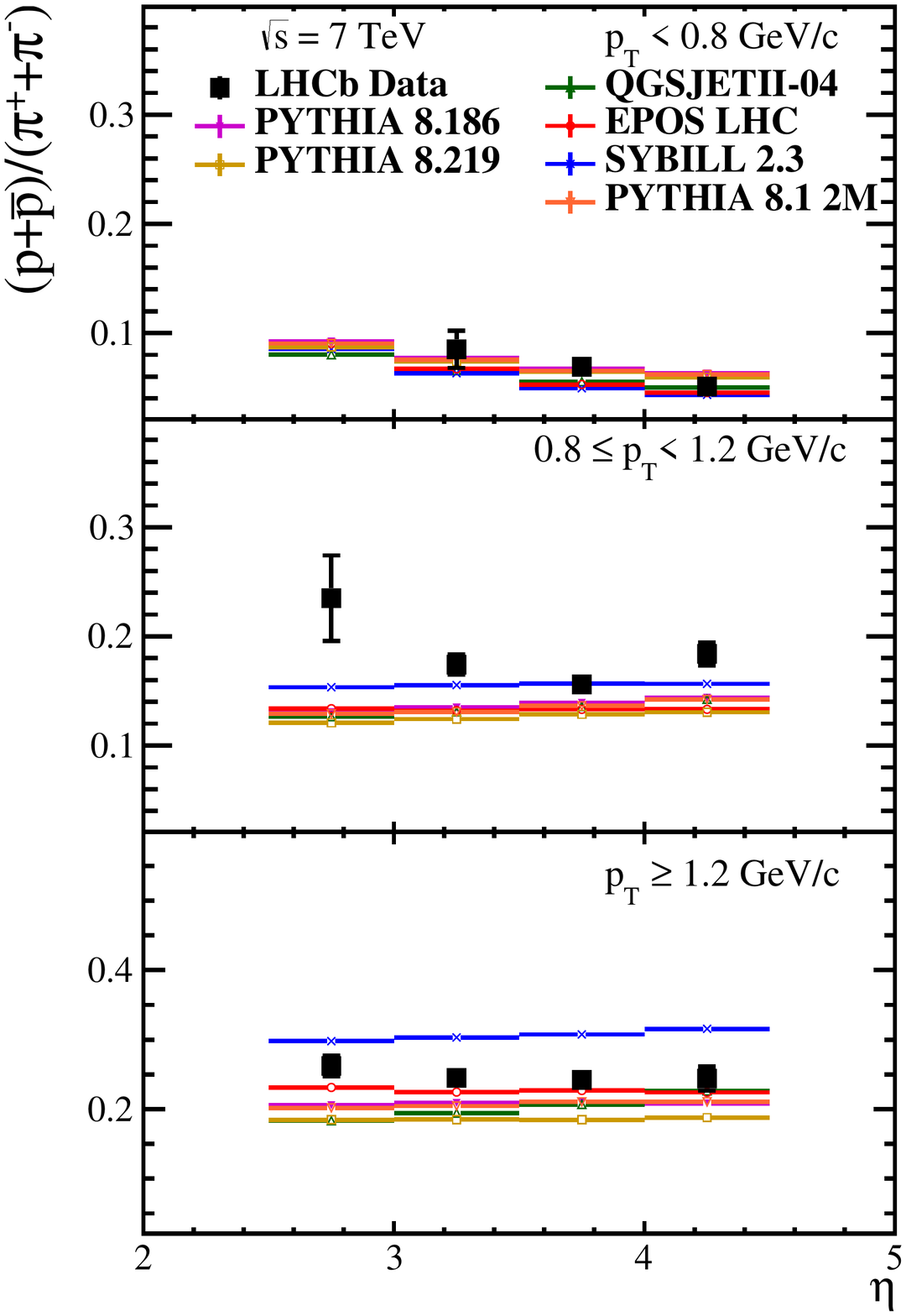} 
  \end{minipage}%%
   \begin{minipage}[b]{0.5\linewidth} 
    \centering
    \includegraphics[trim={0.9cm 0.8cm 0.9cm 1cm},clip,width=0.73\textwidth]{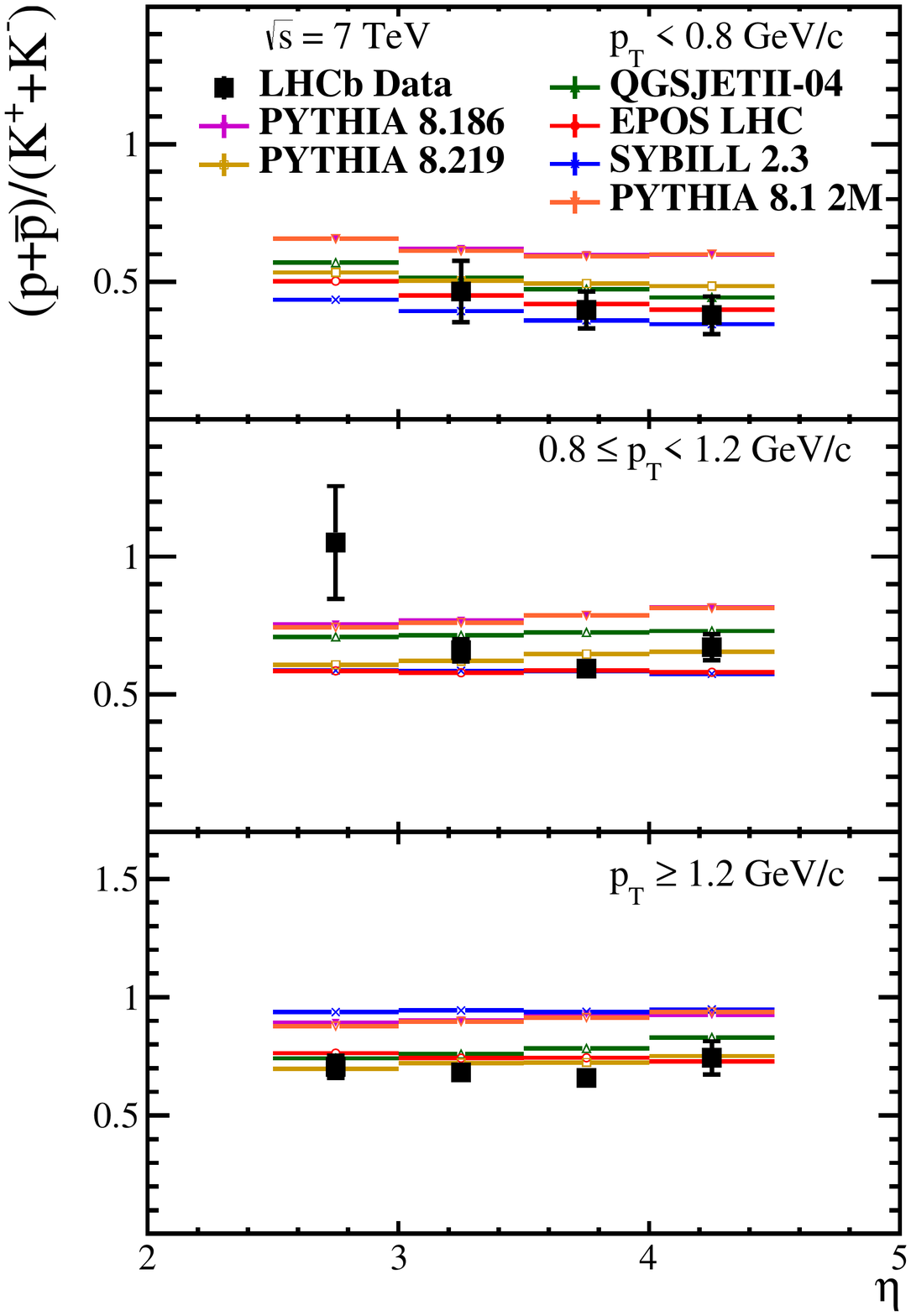} 
  \end{minipage}%%

  \caption{Prompt charged-hadron ratios as a function of pseudorapidity in the kinematic region of $2.5\le\eta\le4.5$ and $p\ge5$ GeV/c in various $p_{T}$  intervals at $\sqrt{s}$ = 7 TeV. The LHCb data vertical bars represent the combined statistical and systematic uncertainties \cite{12}.}

\label{f9}

\end{figure}

A clustering of the predictions in the low $p_{T}$ plot for the $(p+\bar{p})/(\pi^{+}+\pi^{-})$ (shown in Figure \ref{f9}) is observed. Here, all the generators have a good description of the measurements. For the high $p_{T}$ range the closest predictions are the ones of EPOS and PYTHIA 8.1, while for the middle $p_{T}$ range no generator seems to correctly describe the ratio. In the high $p_{T}$ range the ratio is again underestimated by QGSJET, the prediction of which again having an ascending trend, and SIBYLL largely overestimates the ratio.

The $(p+\bar{p})/(K^{+}+K^{-})$ ratio is shown in the same figure. The best prediction overall is the one of EPOS LHC. SIBYLL and QGSJET have a good description of this ratio in the low $p_{T}$ range. In the middle $p_{T}$ range SIBYLL's prediction overlaps with the one of EPOS LHC. In the high $p_{T}$ range PYTHIA 8.219 and QGSJET also have a reasonably good description, although QGSJET exhibits again an ascending trend. SIBYLL again largely overestimates the ratio in this range together with PYTHIA 8.1. The predictions of PYTHIA for the proton/kaon and kaon/pion ratios are clearly improved by the strangeness enhancement from the Monash 2013 tune.

In Figure \ref{fa2} the yields of protons and pions from the high $p_{T}$ region obtained with QGSJET are shown. It is rather clear that the slope of the decrease towards high pseudorapidity of the pions is higher than the corresponding one for the protons. The yields of protons, pions and kaons in the same $p_{T}$ region for all generators are shown in Figure \ref{fa3}. It can be seen that the slope of the proton yield distribution of QGSJET is the lowest, while the one of the pion yield is the highest. The slope of the kaon yield is in between the slopes of the other generators. These together with the observed ascending trend of the QGSJET predictions for the proton/pion, kaon/pion and proton/kaon ratios in the high $p_{T}$ range, while the data or the predictions of the other generators do not show such a trend, suggest that the proton multiplicity decreases too slowly and the pion multiplicity decreases too fast towards high pseudorapidity.

\begin{figure}[h]

  \begin{minipage}[b]{0.5\linewidth} 
    \centering
    \includegraphics[trim={0.9cm 0.8cm 0.7cm 1cm},clip,width=0.73\textwidth]{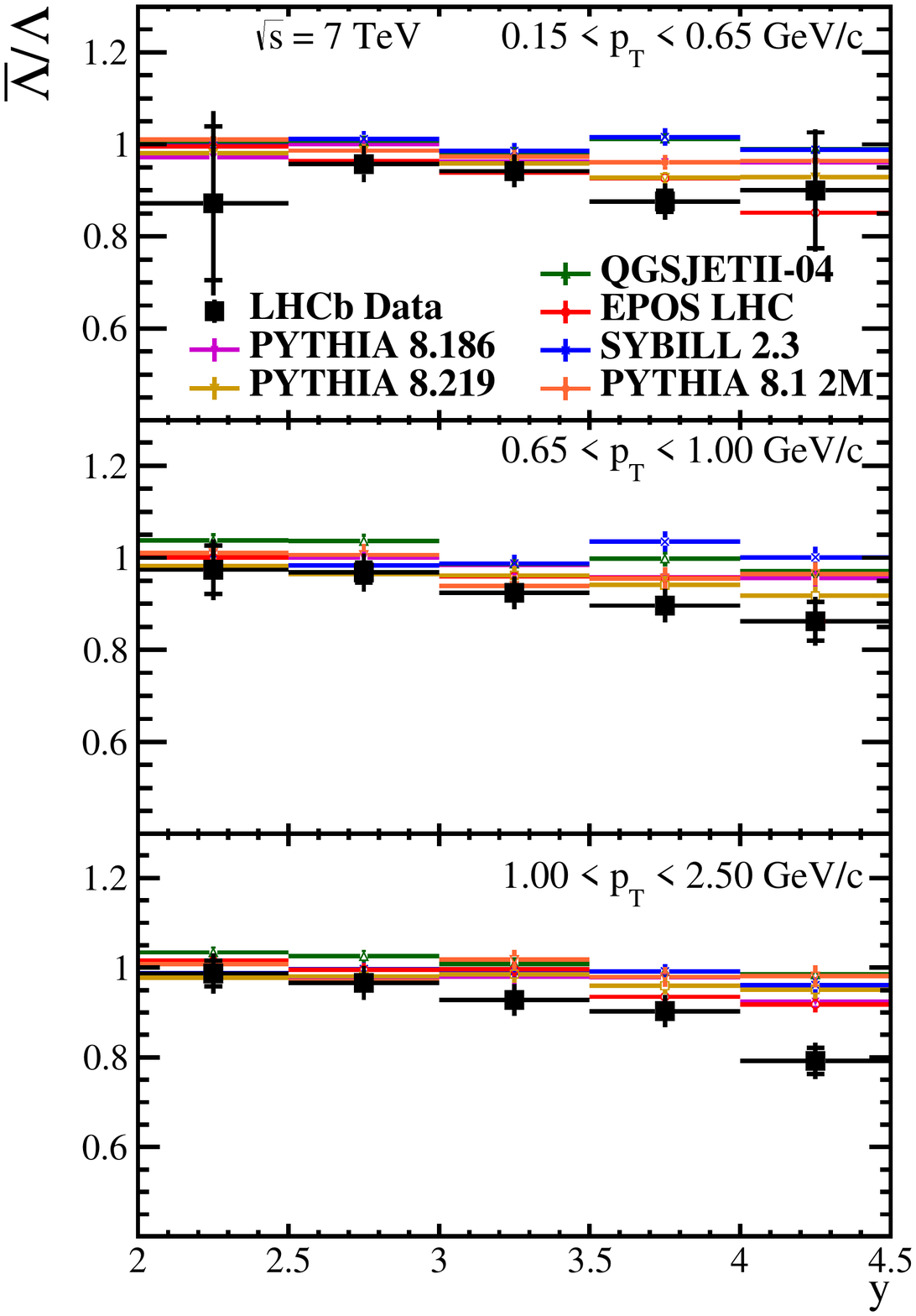} 
  \end{minipage}%%
  \begin{minipage}[b]{0.5\linewidth} 
    \centering
    \includegraphics[trim={0.9cm 0.8cm 0.7cm 1cm},clip,width=0.73\textwidth]{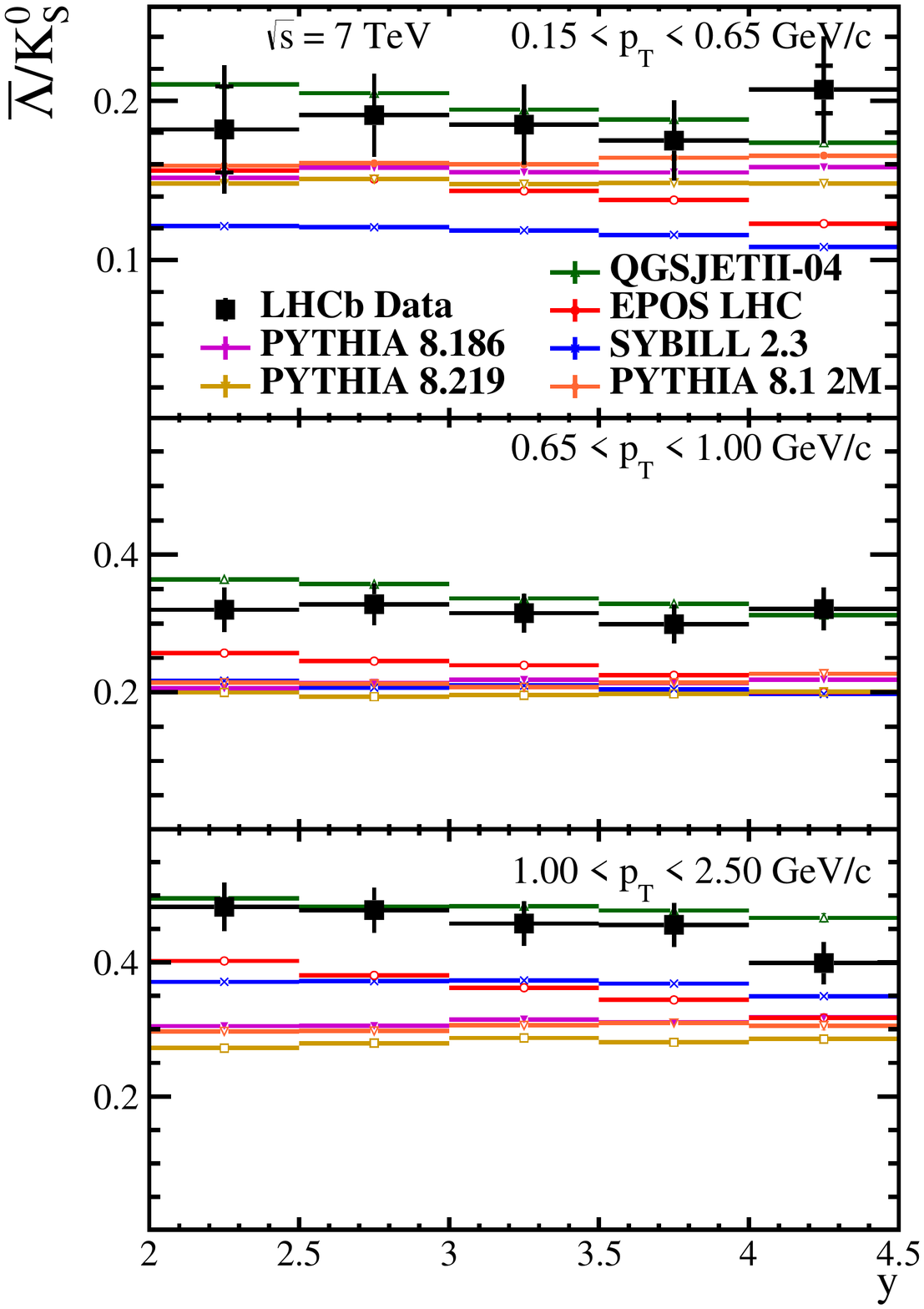} 
  \end{minipage}%%

  \caption{Prompt $V^{0}$ particle ratios as a function of rapidity in the kinematic region of $2\le y \le4.5 $ in various $p_{T}$  intervals at $\sqrt{s}$ = 7 TeV. The LHCb data vertical bars represent the combined statistical and systematic uncertainties and the small horizontal ones show the statistical component \cite{13}.}

\label{f10}

\end{figure}

\begin{figure}[h]

  \begin{minipage}[b]{0.5\linewidth} 
    \centering
    \includegraphics[trim={1.cm 2.cm 0.6cm 3cm},clip,width=0.9\textwidth]{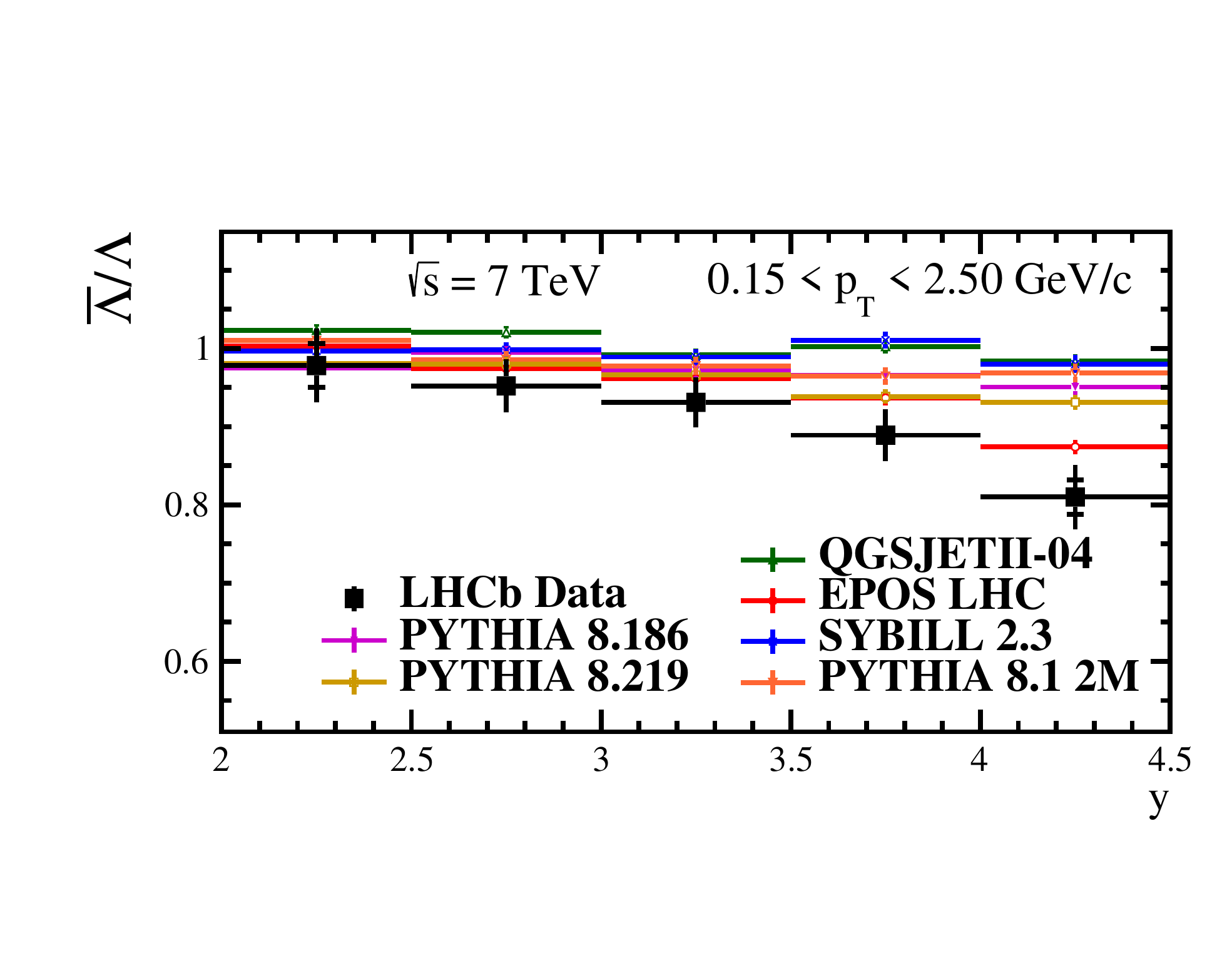} 
  \end{minipage}%%
  \begin{minipage}[b]{0.5\linewidth} 
    \centering
    \includegraphics[trim={1.cm 2.cm 0.6cm 3cm},clip,width=0.9\textwidth]{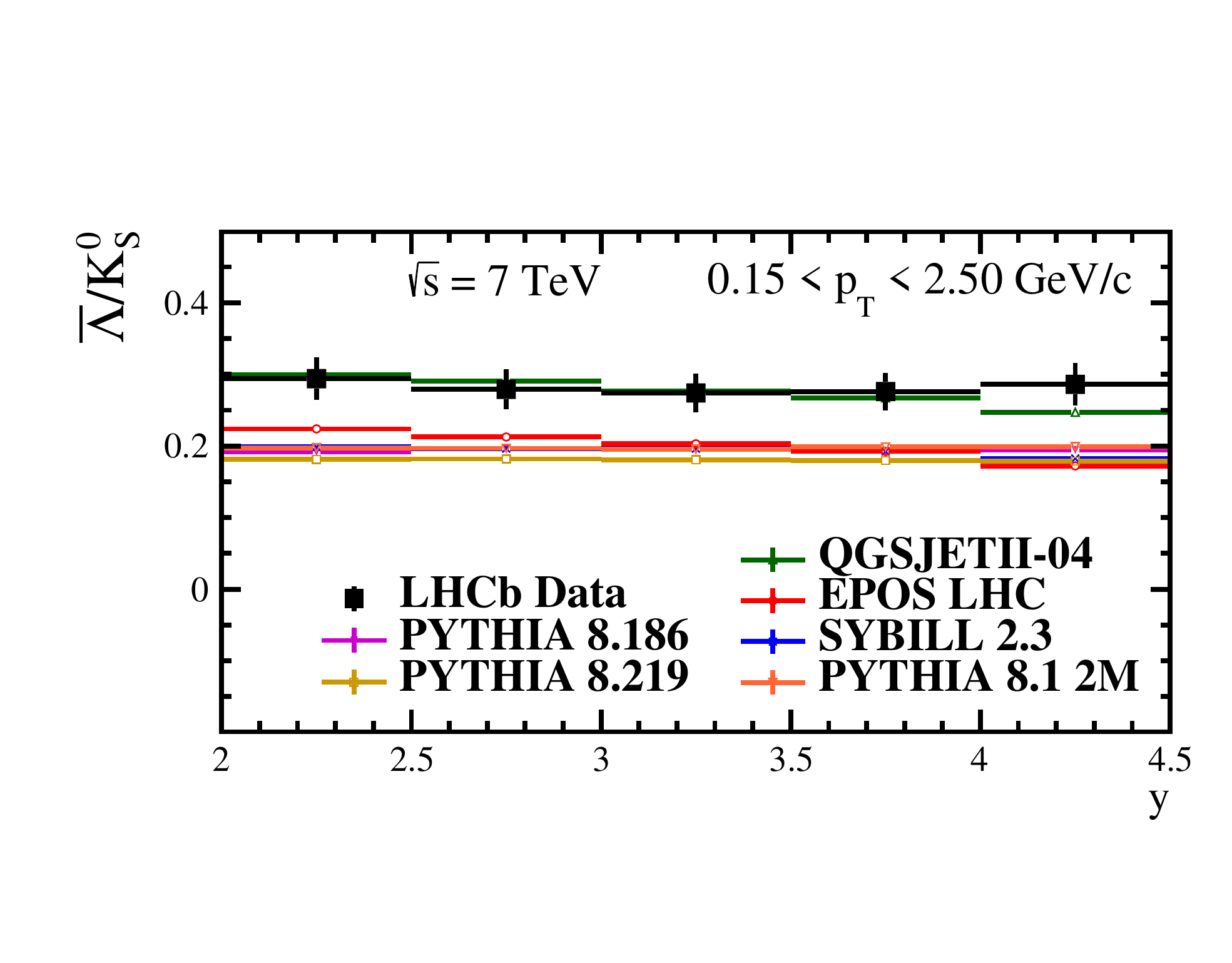} 
  \end{minipage}%%

  \caption{Prompt $V^{0}$ particle ratios as a function of $y$ in the kinematic region of $2\le y \le4.5$ and $0.15<p_{T}<2.50$ GeV/c at $\sqrt{s}$ = 7 TeV. The LHCb data vertical bars represent the combined statistical and systematic uncertainties and the small horizontal ones show the statistical component \cite{13}.}

\label{f11}

\end{figure}

\begin{figure}[h]

  \begin{minipage}[b]{0.5\linewidth} 
    \centering
    \includegraphics[trim={1.cm 2.cm 0.6cm 3cm},clip,width=0.9\textwidth]{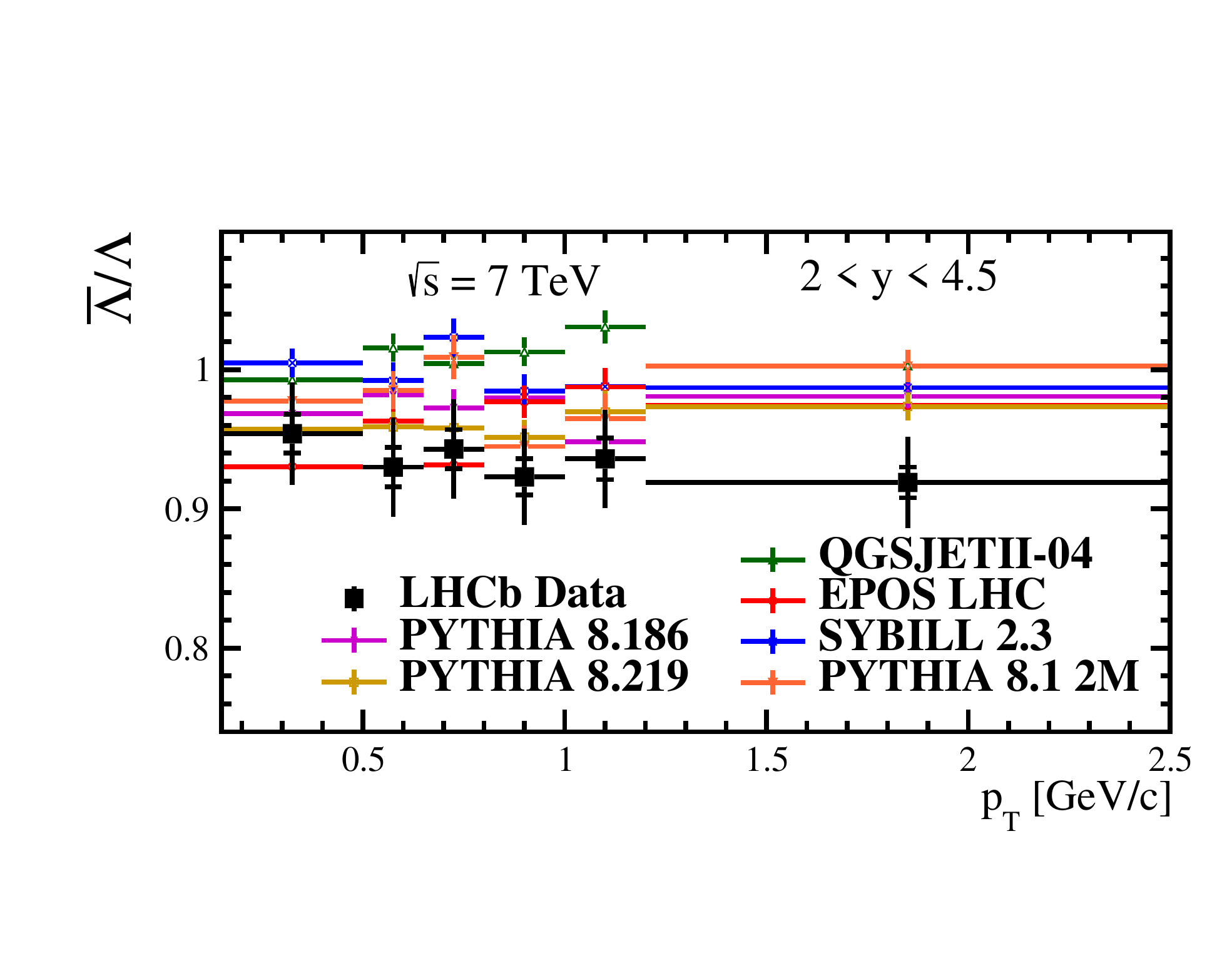} 
  \end{minipage}%%
  \begin{minipage}[b]{0.5\linewidth} 
    \centering
    \includegraphics[trim={1.cm 2.cm 0.6cm 3cm},clip,width=0.9\textwidth]{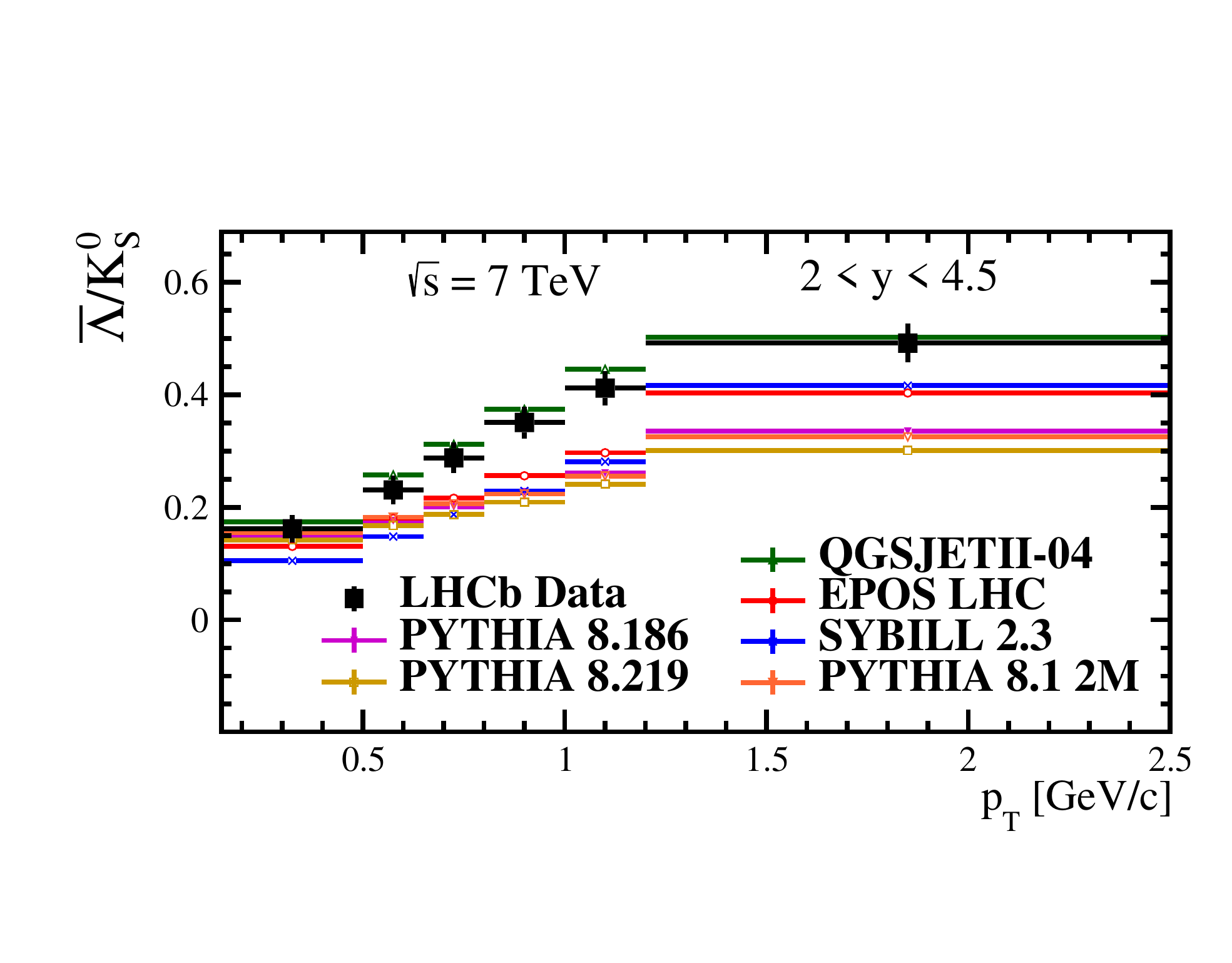} 
  \end{minipage}%%

  \caption{Prompt $V^{0}$ particle ratios as a function of $p_{T}$ in the kinematic region of $2\le y \le4.5$ and $0.15<p_{T}<2.50$ GeV/c at $\sqrt{s}$ = 7 TeV. The LHCb data vertical bars represent the combined statistical and systematic uncertainties and the small horizontal ones show the statistical component \cite{13}.}

\label{f12}

\end{figure}

As one can see in Figures \ref{f10}-\ref{f12}, the $\bar{\Lambda}/\Lambda$ ratio is best described by EPOS LHC and PYTHIA 8.219, pointing to a good baryon number transport. Nonetheless, all predictions have more or less the same trend. The $\bar{\Lambda}/K_{S}^{0}$ ratio seems to be reasonably well described by QGSJET, while the other generators largely underestimate it.

\begin{center}
    \includegraphics[trim={0.7cm 0.4cm 0.7cm 0.8cm},clip,width=0.35\textwidth]{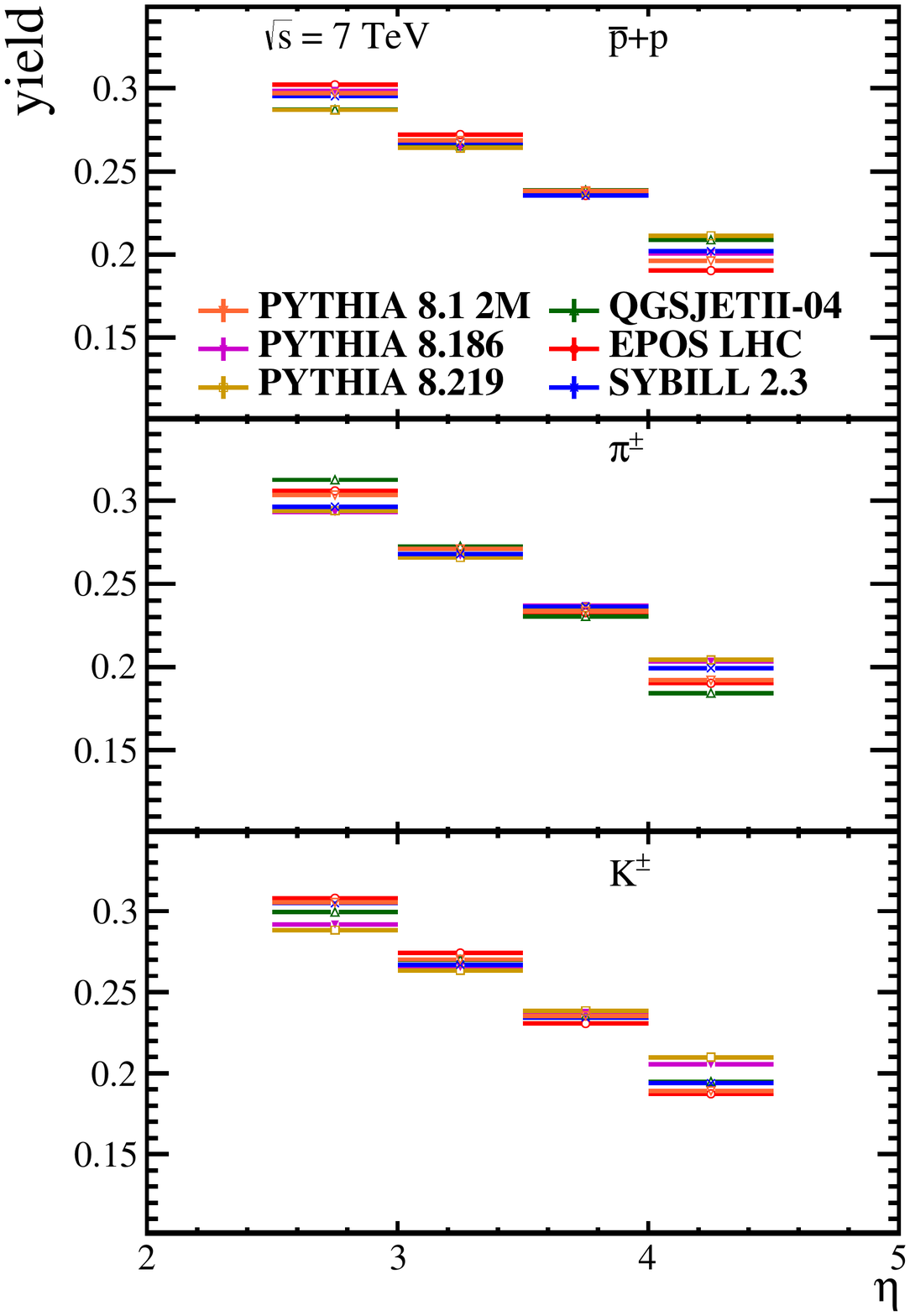}0
 \figcaption{\label{fa3} Yields (normalized to 1) of protons, pions and kaons with $p_{T} \geq 1.2$ GeV/c and $p \geq$ 5 GeV/c.}

\end{center}

\begin{center}
    \includegraphics[trim={0.7cm 0.4cm 0.7cm 0.8cm},clip,width=0.4\textwidth]{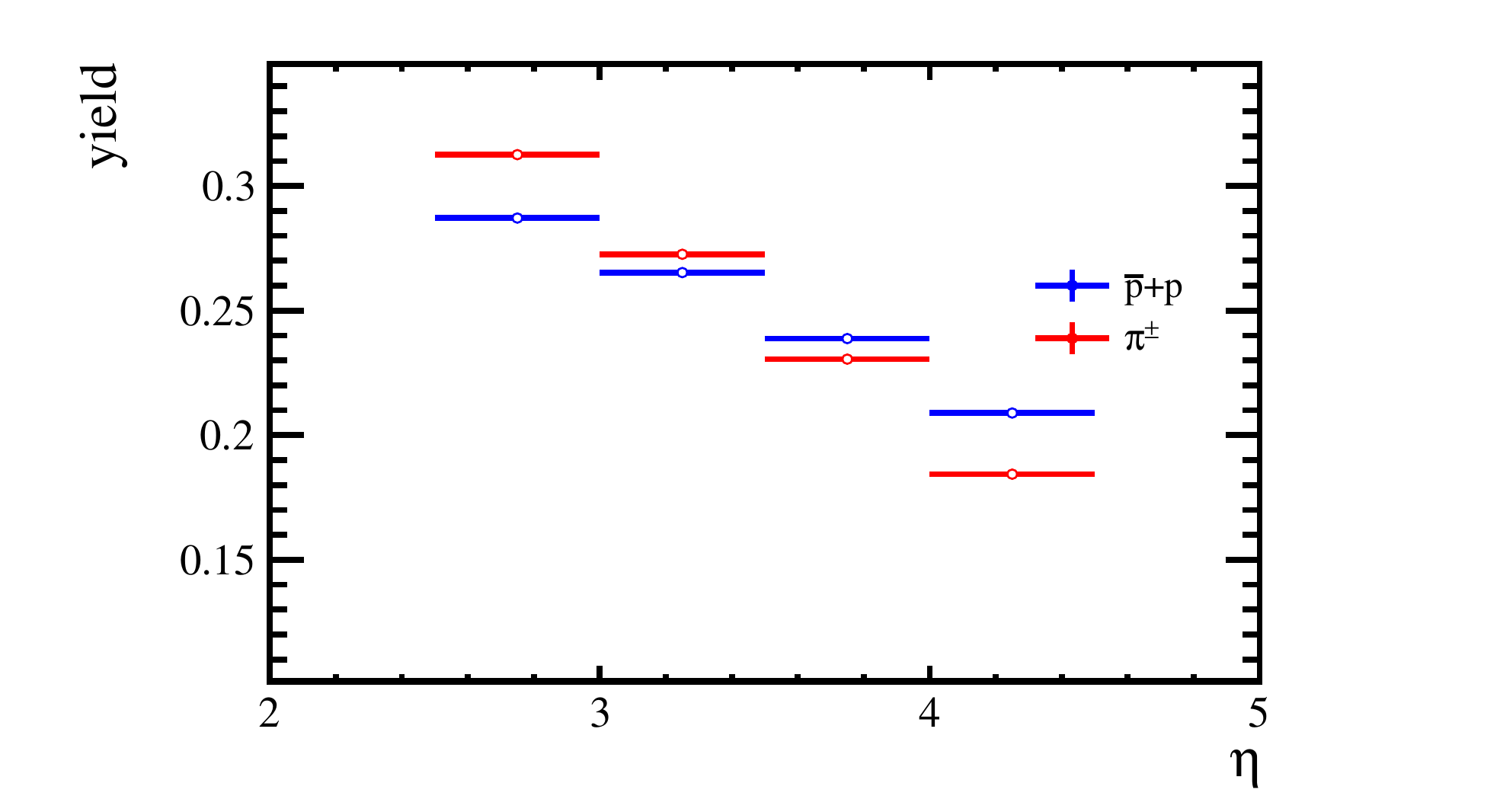}
 \figcaption{\label{fa2} Yields (normalized to 1) of protons and pions with $p_{T} \geq 1.2$ GeV/c and $p \geq$ 5 GeV/c generated with QGSJETII-04.}

\end{center}

\clearpage
\end{multicols}

%Ratios1...............

%Ratios3...............

\clearpage

%Ratios5...........

%Ratios7...........

\clearpage

%Ratios9...........

%Ratios11...........

\begin{multicols}{2}

%.............................................

\end{multicols}
\begin{multicols}{2}

\section{Conclusions}

The generators that have been studied are EPOS LHC, QGSJETII-04, SIBYLL 2.3 and versions 8.186 and 8.219 of PYTHIA. The observables on which the study was conducted were the charged energy flow, charged-particle multiplicities and densities, charged-hadron production ratios, $V^{0}$ ratios and other strange particle distributions. It is reasonably clear that no generator reproduces the data for all of the observables studied, but rather one generator describes well only a particular set of the observables or aspects of particle production. As a general trend, the predictions are better in the central region. The tuning using data from the central-rapidity range of general purpose LHC detectors is visible and clearly improves the estimations even for the forward region, though the effect of extrapolation to higher rapidity is in clear disagreement with experimental data.

It was observed that the charged energy flow, which can be regarded as a global event observable, is relatively well described by all the generators, at least in terms of shape. The best prediction overall for the charged energy flow is that of SIBYLL 2.3, a generator tuned specifically to reproduce correctly this type of observable. PYTHIA 8.186 has the best description of the other LHC-tuned generators.

EPOS and PYTHIA, especially version 8.219, are very similar in their description of the observables. The similarity between the generators may arise from the partonic approach and similar perturbative calculations that they both use for hard parton collisions.

QGSJET is similar to EPOS in the description of some observables like the charged energy flow (except for the hard event class) and charged particle densities, but also shares some similarities with SIBYLL.

The multiplicity distributions are generally not well reproduced by the generators. Here EPOS and PYTHIA have the best predictions overall. Also, they seem to get better with the increasing hardness of the processes, but exhibit a similar effect to the one of the other generators, i.e., favouring either very low or high multiplicity events, albeit at a much lower level than SIBYLL, for example, which has the most polarizing behaviour.

SIBYLL has a few notable successes in describing some particle ratios and also its predictions for charged particle pseudorapidity and transverse momentum distributions have a good shape.

The best baryon transport mechanism seems to be the one of EPOS, followed by the one of PYTHIA, while the $\bar{\Lambda}/K_{S}^{0}$ ratio is best described by QGSJET.

Most of the observed differences seem to be an effect of extrapolation in the forward region. So, the extrapolation uncertainties seem to be rather large. Nonetheless, in the majority of cases, the measurements fall within a band defined by the most extreme predictions.

The relative contributions of particle production processes differ between the central and forward regions. In the central pseudorapidity region there is a significant contribution of hard parton-parton scatterings (with high squared momentum transfer) to which high multiplicity events and high $p_{T}$ jets are associated. In the forward region, on the other hand, the underlying event (multiparton interactions and beam remnants), as well as diffractive processes have a considerable contribution. The event generators usually have different sets of parameters for each process and as such, when tuning using measurements from one pseudorapidity region or the other, different parameters are constrained, so each tune is applicable for studies in its respective region. As shown in this paper, the predictions in the forward region are improved by the tuning of the generators using measurements from the central region, but it seems that a dedicated tuning procedure is still necessary. So, the utility of each tune is somewhat limited when extrapolating from the central to the forward region and vice versa. Ideally, measurements from both the forward and central regions should be used simultaneously when tuning a generator, but this is seldomly happening. In many cases there are intrinsic limitations of the generators or the models they are based on, which prevent a simultaneous tune in both regions and so, a more consistent overall description of the processes. Difficulties related to such a tuning procedure also arise from the different experimental conditions in each region.

As we have seen in this paper, it seems that the modelling of the soft processes is still open to improvement and a forward tuning of generators is required to improve precision in this rapidity range. Hence, it may prove useful to take into account during the tuning process measurements from LHCb and TOTEM, which are LHC experiments in the forward region, where the soft process component is sensibly larger than in the central region, the baryon transport is different, and the multi parton collisions might give a different signal.

\acknowledgments{We would like to thank the authors of PYTHIA/EPOS/QGSJET/SIBYLL generators and the authors of the CRMC interface (T. Pierog, C. Baus, and R. Ulrich).}

\end{multicols}

\vspace{-1mm}
\centerline{\rule{80mm}{0.1pt}}
\vspace{2mm}

\begin{multicols}{2}

\end{multicols}

\end{document}